\documentclass[12pt,preprint]{aastex}
\usepackage{pdflscape}
\usepackage[titletoc,toc,title]{appendix}

\shorttitle{GEMINI observations of NGC 6624}
\shortauthors{Saracino et al.}

\begin{document}

\title{Ultra-deep GEMINI near-infrared observations of the bulge
  globular cluster NGC 6624\footnote{\footnotesize Based on
    observations obtained at the Gemini Observatory, which is operated
    by the Association of Universities for Research in Astronomy,
    Inc., under a cooperative agreement with the NSF on behalf of the
    Gemini partnership: the National Science Foundation (United
    States), the National Research Council (Canada), CONICYT (Chile),
    the Australian Research Council (Australia), Minist\'{e}rio da
    Ci\^{e}ncia, Tecnologia e Inova\c{c}\~{a}o (Brazil) and Ministerio
    de Ciencia, Tecnolog\'{i}a e Innovaci\'{o}n Productiva
    (Argentina).  Based on observations gathered with ESO-VISTA
    telescope (program ID 179.B-2002).}  }

\author{S. Saracino\altaffilmark{1,2},
  E. Dalessandro\altaffilmark{1,2}, F. R. Ferraro\altaffilmark{1},
  D. Geisler\altaffilmark{3}, F. Mauro\altaffilmark{4,3},
  B. Lanzoni\altaffilmark{1}, L. Origlia\altaffilmark{2},
  P. Miocchi\altaffilmark{1}, R. E. Cohen\altaffilmark{3}, S. Villanova\altaffilmark{3}, C. Moni
  Bidin\altaffilmark{5} }

\altaffiltext{1}{Dipartimento di Fisica e Astronomia, Universit\`a di
  Bologna, Viale Berti Pichat 6/2, I-40127 Bologna, Italy  -  \it {sara.saracino@unibo.it}}
\altaffiltext{2}{INAF - Osservatorio Astronomico di Bologna, via
  Ranzani 1, I-40127 Bologna, Italy}
\altaffiltext{3}{Departamento de Astronom\'ia, Universidad de
  Concepci\'on, Casilla 160-C, Concepci\'on, Chile}
\altaffiltext{4}{Millennium Institute of Astrophysics, Chile}
\altaffiltext{5}{Instituto de Astronom\'ia, Universidad Cat\'olica del
  Norte, Av. Angamos 0610, Antofagasta, Chile}

\begin{abstract}
We used ultra-deep $J$ and $K_s$ images secured with the near-infrared
GSAOI camera assisted by the multi-conjugate adaptive optics system
GeMS at the GEMINI South Telescope in Chile, to obtain a ($K_s$,
$J-K_s$) color-magnitude diagram (CMD) for the bulge globular cluster
NGC 6624. We obtained the deepest and most
accurate near-infrared CMD from the ground
for this cluster, by reaching $K_s$ $\sim$ 21.5, approximately 8 magnitudes
below the horizontal branch level. The entire extension of the
Main Sequence (MS) is nicely sampled and at $K_s$ $\sim$ 20 we
detected the so-called MS ``knee'' in a purely near-infrared
CMD. By taking advantage of the exquisite quality of the data, we
estimated the absolute age of NGC 6624 ($t_{age}$ = 12.0 $\pm$ 0.5
Gyr), which turns out to be in good agreement with previous studies in
the literature. We also analyzed the luminosity and mass functions of
MS stars down to M $\sim$ 0.45 M$_{\odot}$ finding evidence of a
significant increase of low-mass stars at increasing distances from
the cluster center.  This is a clear signature of mass segregation,
confirming that NGC 6624 is in an advanced stage of dynamical
evolution.
\end{abstract}

\date{September 7, 2016}


\section{Introduction}
Globular Clusters (GCs) are complex systems hosting $10^{4}-10^{6}$
gravitationally bound stars, distributed with an approximately
spherical geometry.  In the central regions of these stellar systems,
where stars are forced to live in an extremely dense environment, the
probability of stellar encounters is highly enhanced. Such collisions
lead to the formation of peculiar objects like Cataclysmic Variables,
Low Mass X-ray Binaries (LMXBs), Millisecond Pulsars and Blue
Straggler Stars \citep[e.g.][]{Ba92,Pa92,Fe01,Fe09a,Fe12,Ra05,PH06}
and they influence the time scales on which mass segregation,
core collapse and other dynamical processes occur \citep{Mey97}.

The large stellar concentration in the central regions of high density
GCs prohibited to resolve individual stars
for many years, until the launch of Hubble Space Telescope (HST),
which allowed exquisite resolution over a relatively large (a few
square arcmin) field of view (FOV) mainly in the optical bands. The
investigation of the stellar content of GCs in the bulge of the Galaxy
has the additional complication of the presence of thick clouds of dust along the line of sight 
that almost totally absorb the optical light, and/or the presence of a heavy field star contamination. The exploration of such
clusters requires near-infrared (NIR) observations
\citep[e.g.,][]{Fe00,Fe09b,Va04a,Val07,Or97,Or11}, a wavelength range
where the foreground extinction significantly drops. The observations of such 
stellar systems in 
the NIR have tremendously improved with the advent of 
instrumentation assisted by the new adaptive optics (AO) facilities. These
systems, in fact, mainly work in the NIR, where the spatial and
temporal coherence of the corrugated wavefront is larger \citep{DK12},
and they are able to compensate the blurriness of the astronomical
images due to the Earth's atmosphere by using one or more deformable mirrors and,
as reference, natural and/or laser guide stars.

The Gemini South Telescope, on Cerro Pach\'on in Chile, currently is
the only facility equipped with a Multi-conjugate Adaptive Optics
System (named GeMS), which uses three natural guide stars, a
constellation of five laser guide stars and two deformable mirrors
conjugated at the ground and at an altitude of 9 km
\citep[see][]{Rig14,Nei14}. This allows GeMS to (almost) reach the diffraction
limit of the telescope.  By exploiting the unprecedented capabilities
of this system, which works in combination with the NIR
high-resolution camera Gemini South Adaptive Optics Imager (GSAOI), we
started an observing campaign of a set of GCs located in the Galactic
bulge. After the encouraging results obtained for Liller 1
\citep{Sar15}, a heavily obscured cluster located very close to the
Galactic plane and center, here we present the results for NGC 6624. This GC is
located just at the edge of the inner bulge, at a distance of 7.9 Kpc from Earth (\citealp{Har96}, 2010 edition) and it is characterized by only a moderate foreground 
extinction for a bulge GC, $E(B-V) = 0.28$
(\citealp{Va04a, Har96}, 2010 edition). 
These features also make NGC 6624
an ideal target to investigate the sky performance of the
GeMS+GSAOI system.

NGC~6624 is a well-studied cluster, but mainly in the optical bands
\citep[see][]{Sar07,Dal14}. In the NIR it was observed with the following instruments:
\begin{itemize}
\item[{\it (i)}] IRCAM, mounted at the 2.5 m Du Pont
  telescope. These observations provided the first ($K_s$, $J-K_s$)
  CMD of the cluster sampling the brightest portion of the red giant
  branch (RGB; see \citealp{KF95}) down to the Horizontal Branch (HB)
  level.
\item[{\it (ii)}]  IRAC-2 mounted at the ESO 2.2 m telescope
  MPI. The NIR CMD derived from these observations was deeper than the
  previous one, reaching the cluster sub-giant branch ($K_s$ $\sim$
  17), and thus allowing to study the RGB features (RGB bump and
  tip; see \citealp{Va04a, Va04b,Val07,Fer06b}).
\item[{\it (iii)}] Within the {\it VISTA Variables in the Via Lactea}
  (hereafter VVV) survey.  The NIR CMD obtained from the VVV catalog
  was deep enough ($K_s$ $\sim$ 19) to sample the main sequence
  turn-off (MS-TO) only in the external regions of the cluster
  \citep{Min10,Cat11}.
\end{itemize}
NGC 6624 is quite compact and it has been catalogued as a dynamically
evolved cluster, which already experienced core collapse
\citep{Tra95}. It has been found to harbour six millisecond pulsars
\citep[][]{Lyn12,Tam11,Fre11}
and at least an ultra-compact LMXB \citep{Dib05}, thus confirming that its dense environment
efficiently boosts the formation of exotic objects.

In this paper we present ultra-deep NIR observations of NGC 6624
obtained by using the powerful combination of the GeMS+GSAOI devices
mounted at the Gemini South Telescope. The paper is organized as
follows: in Section \ref{obs} we discuss the observations and the data
analysis. In Section \ref{cmd} we present the CMD of the
cluster. Section \ref{age} is focused on the determination of the age
of the cluster and in Section \ref{luminosity} we discuss the
Luminosity (LF) and Mass Functions (MF) of the cluster MS. In
Section \ref{concl} we present our conclusions.

\section{Observations and data analysis}
\label{obs}
\begin{figure}
\epsscale{1.}
\plotone{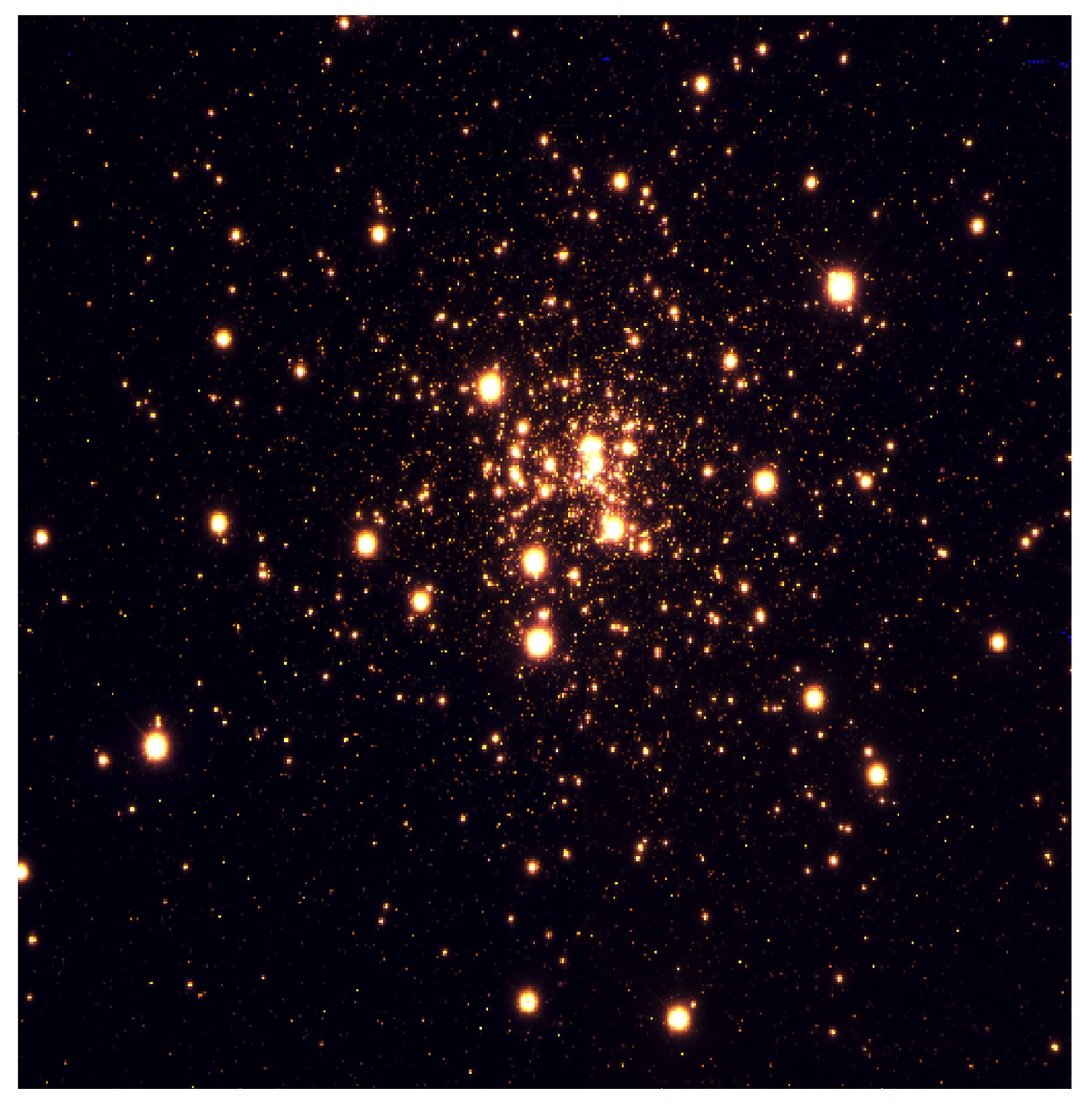}
\caption{Two-color image of NGC 6624 obtained by combining GEMINI
  observations in the NIR $J$ and $K_s$ bands. North is up, east is on
  the left. The field of view is $93\arcsec\times 93\arcsec$.}
\label{fig1}
\end{figure}

\begin{figure}
\epsscale{1.}
\plotone{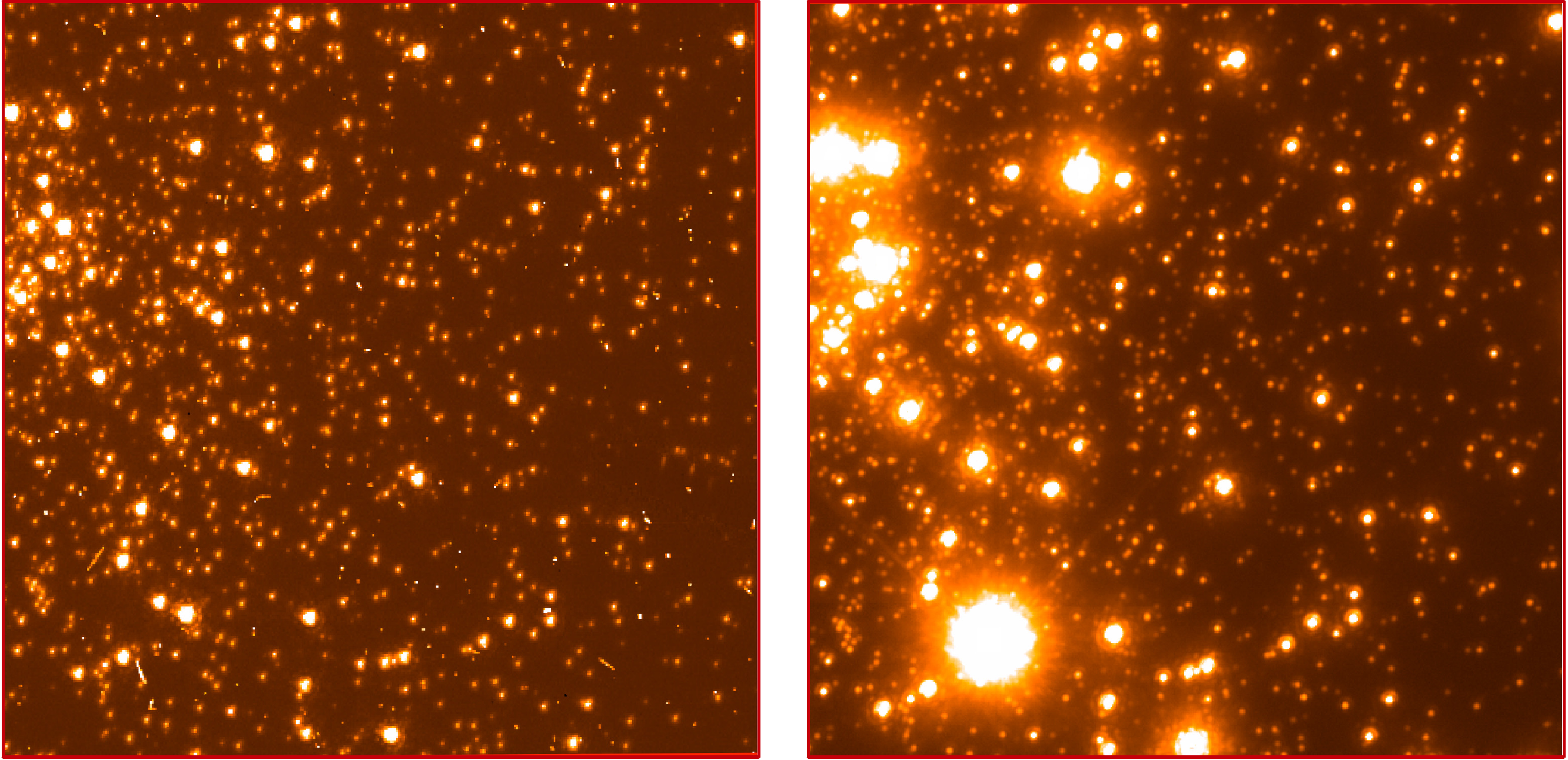}
\caption{A zoom-in of a central region ($10\arcsec\times10\arcsec$) of
  the cluster as seen by ACS/HRC (F435W filter) on board HST
  (\citealp{Dal14}, {\it left}) and by GSAOI+GeMS in the $K_s$ band
  (this work, {\it right}). The spatial resolution of GEMINI in the
  NIR turns out to be comparable to that of HST in the optical.}
\label{zoomGEMS}
\end{figure}

The data set analyzed in this work consists of a sample of
high-quality $J$ and $K_s$ images obtained with GeMS and the
NIR camera GSAOI. GSAOI has a resolution of about 0.02\arcsec/pixels
and consists of four 2048 x 2048 pixels chips, divided by gaps of
about 2\arcsec, providing a total FOV of almost $85\arcsec \times
85\arcsec$. NGC 6624 was observed on May 24, 2013 as part of the
proposal GS-2013-Q-23 (PI: D. Geisler).  A total of 28 images (14 in
the $J$ and 14 in the $K_s$ bands), with an exposure time $t_{\rm
  exp}= 30$ s each, was acquired during the run. To fully cover the
gaps between the GSAOI chips, we adopted a dither pattern with
a maximum offset of about $5\arcsec$ between two consecutive images.
During the observing night the DIMM monitor on Cerro Pach\'on recorded
excellent seeing conditions, with a Full Width at Half Maximum (FWHM)
of $\sim 0.7\arcsec$. An average stellar FWHM smaller than 4 pixels
(0.08\arcsec, close to the diffraction limit of the telescope) has been measured in all $J$ and $K_s$ images.  
Unfortunately, one $J$-band exposure
was characterized by a significantly worse FWHM, so it was excluded
from the subsequent analysis.  In Figure \ref{fig1} we present a
two-color mosaic of $K_s$ and $J$ images of NGC 6624. Figure \ref{zoomGEMS}
shows the comparison between a central region (10\arcsec $\times$
10\arcsec) of NGC 6624 obtained with HST in the F435W filter ({\it
  left panel}) and with GeMS/GSAOI system in the $K_s$ band ({\it
  right panel}): the spatial resolution of the AO corrected image is indeed 
  impressive.

\begin{figure}
\epsscale{1.}
\plotone{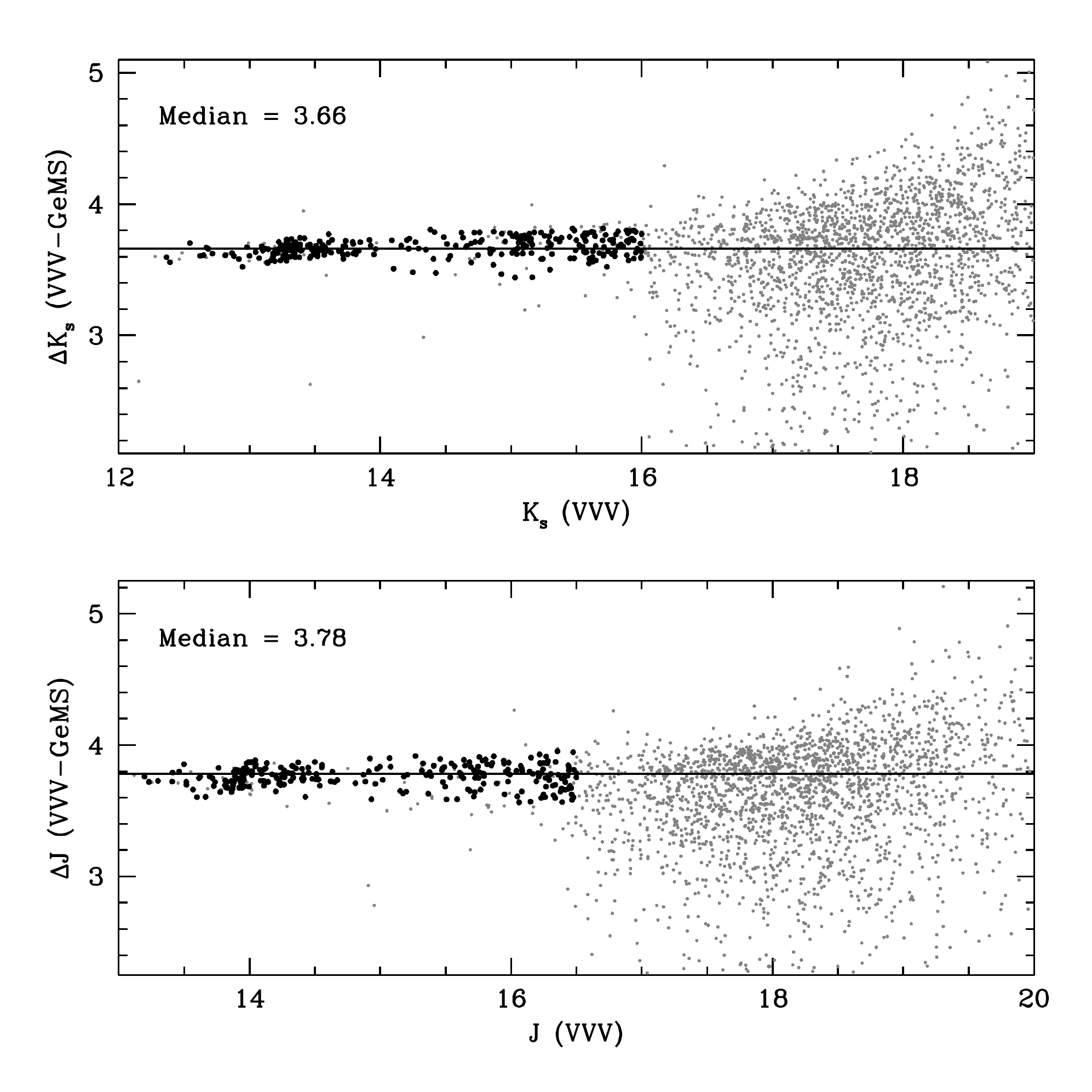}
\caption{Photometric calibration plots for the GEMINI catalog of NGC 6624 in the
  $K_s$ and $J$ bands.  Only bright stars (black points) have been
  used to determine the calibration zero points. The median values,
  estimated by using a 2$\sigma-$rejection, are shown in the
  figure.}
\label{calib}
\end{figure}

The data analysis was performed following the same procedure described
in \citet{Sar15}. Briefly, each individual
chip\footnote{To distinguish between chips, we assigned an
  identification (ID) number from 1 to 4, starting from the
  bottom-right corner (chip 1) and proceeding clockwise.} was analyzed independently, by using 
  standard pre-reduction procedure, within the
\texttt{IRAF}\footnote{IRAF is distributed by the National Optical Astronomy
  Observatory, which is operated by the Association of Universities
  for Research in Astronomy, Inc., under cooperative agreement with
  the National Science Foundation.}  environment, in order
to correct for flat-field and bias and subtract the background. For this purpose, a 
Master Sky was obtained by combining 5 sky images per
filter of a relatively empty field. In order to get
accurate photometry, even in the innermost region of the cluster, we
performed point-spread function (PSF) fitting by using \texttt{DAOPHOT} \citep{Ste87} of all the detected
stellar sources.  For each chip, we selected $\sim$ 100 bright,
isolated and unsaturated stars (using \texttt{FIND} and \texttt{PHOTO}), that were used
to derive the best-fit PSF model. Based on a $\chi^2$ test, the
best-fit PSF analytic models were found to be a MOFFAT function (with
$\beta$=1.5; \citealp{Mof69}) for the $K_s$ images and a PENNY
function \citep{Pen76} for the $J$ ones. We also noted that photometry significantly improves 
by adopting PSF models varying within the FOV. In particular we adopted
a linear dependence on the position within the frame in chip 1, and a cubic one 
for chips 2, 3 and 4.
The selected
models were then applied, using the \texttt{ALLSTAR} and \texttt{ALLFRAME} tasks
\citep{Ste94}, to all the sources having peak counts larger than
3$\sigma$ above the local background level, thus determining the
instrumental magnitudes for each star candidate in each chip. In the
catalog we included only stars present in at least two images for each
filter. This criterion removed most of the cosmic rays and other
spurious detections and allowed to fill the gaps between the different
chips of GSAOI. For each star, we homogenized the magnitudes estimated
in different images, and their weighted mean and standard deviation
have been finally adopted as the star mean magnitude and its related
photometric error \citep[see][]{Fer91,Fer92}. The instrumental
positions have been reported onto the absolute coordinate system by
adopting the star catalog by \citet{Dal14}
as astrometric reference and by using the cross-correlation software
\texttt{CataXcorr} \citep{Mon95}.
The instrumental magnitudes of the final catalog have been finally
calibrated by using a sample of stars in common with the VVV
catalog. The latter has been realized by using the 
\texttt{DAOPHOT} based VVV-SkZ pipeline \citep{Mau13} and reported into the 
2MASS photometric and astrometric system \citep[see details in][]{chene12,monibidin11,Mau13}. We adopted an iterative 
2$\sigma$ - clipping procedure to
estimate the calibration equations transforming the instrumental GeMS
$K_s$ and $J$ magnitudes into the VVV photometric system. In order to
avoid bias, only the brightest stars (with $K_s<16$ and $J<16.5$) were
used (see Figure \ref{calib}), yielding a total of about 300 calibration stars in each filter. The r.m.s. of the calibration is
$\sim$ 0.02 mag both in $J$ and in $K_{s}$.

\section{Near-Infrared Color-Magnitude diagrams of NGC 6624}
\label{cmd}
\begin{figure}
\epsscale{1.}
\plotone{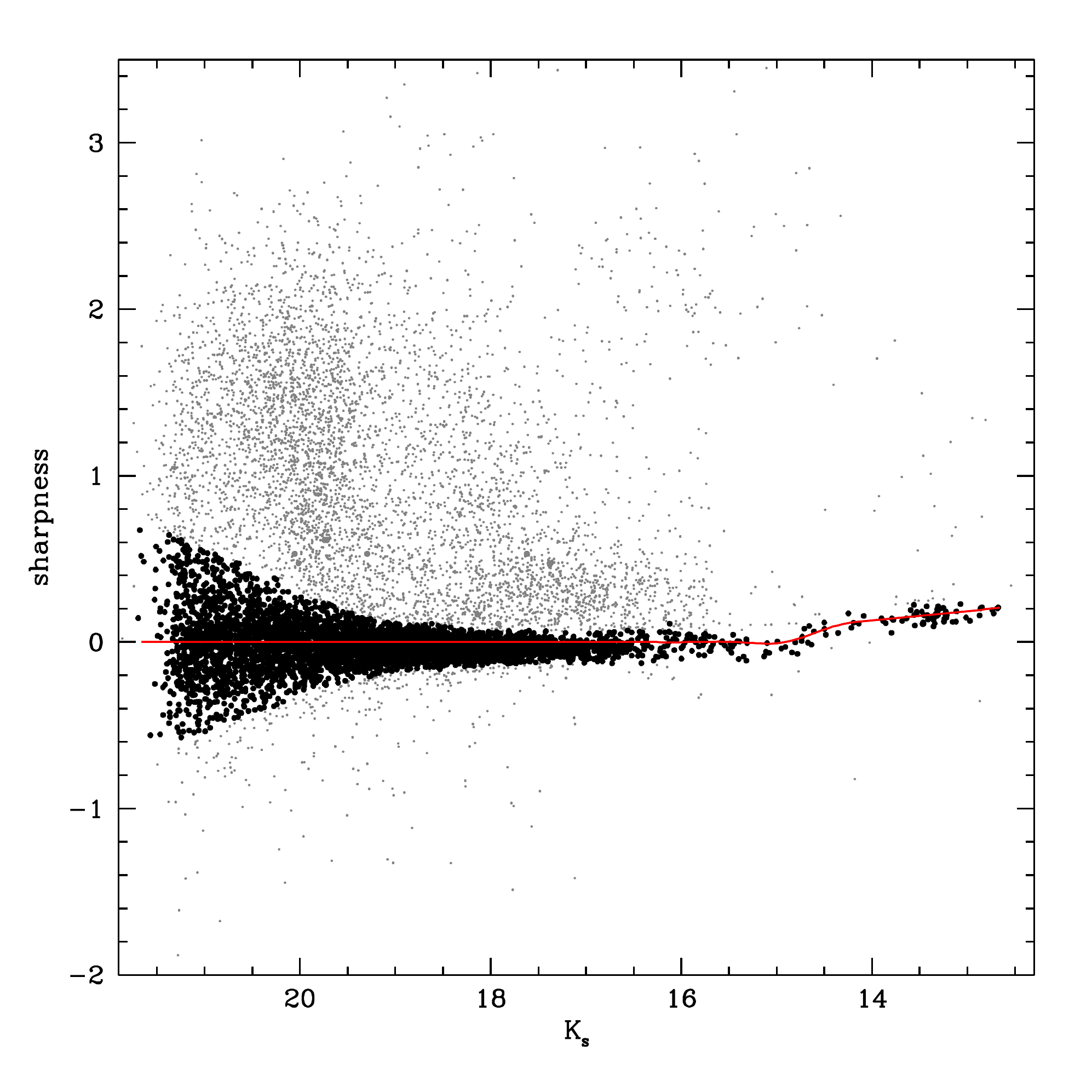}
\caption{Plot of the sharpness as a function of the $K_s$ magnitude.
  The black points are the stars lying within 6$\sigma$ from the red
  line, which represents the 2$\sigma$-clipped median value in
  sharpness. This selection criterion is applied to all the detected
  stars in order to remove spurious objects and stars with large
  photometric errors.}
\label{sharp}
\end{figure}

\begin{figure}
\epsscale{1.}
\plotone{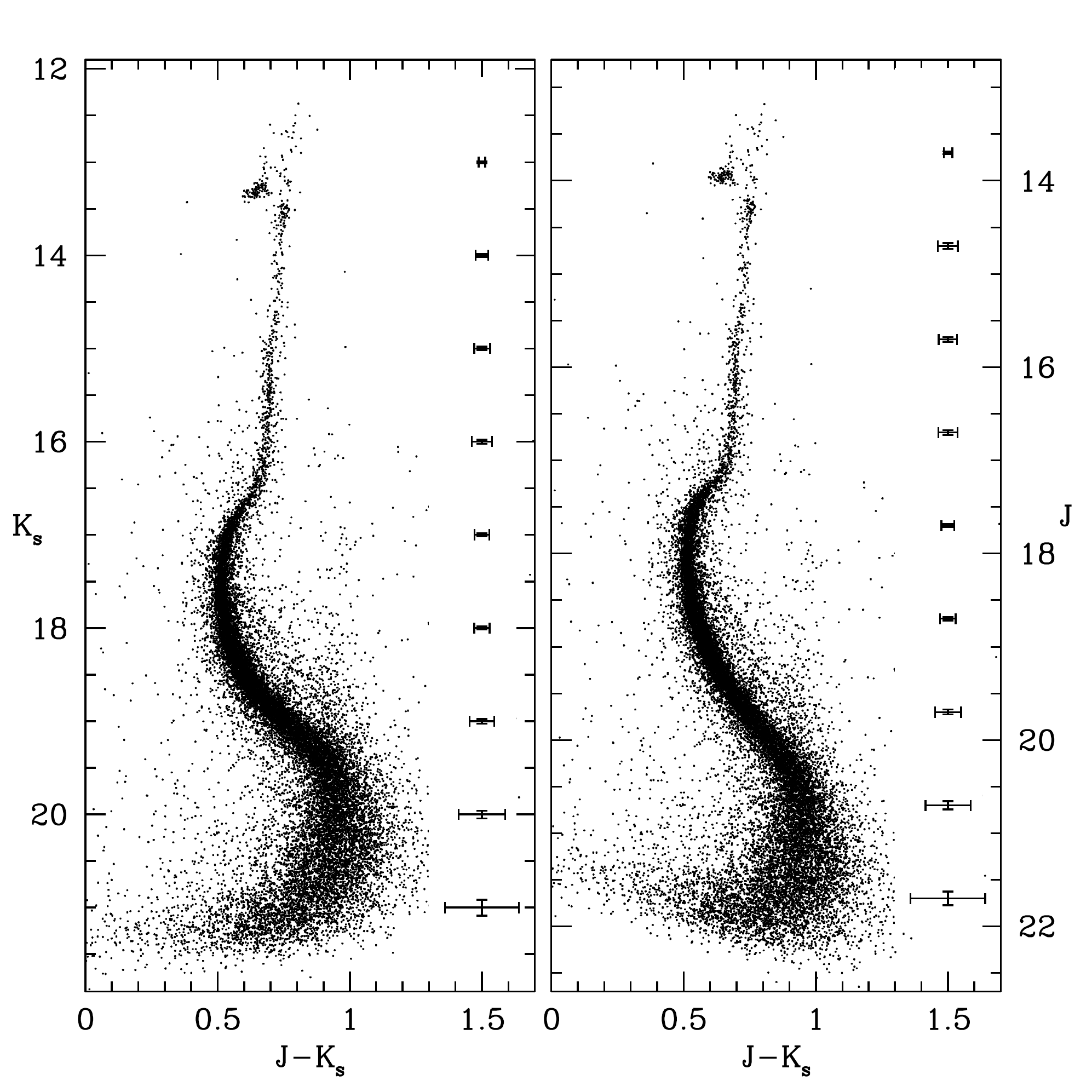}
\caption{($K_s, J-K_s$) and ($J, J-K_s$) CMDs of NGC 6624 obtained
  from the GEMINI observations discussed in the paper.  All the main
  evolutionary sequences of the cluster are well visible, from the
  RGB, HB, MS-TO down to the MS-K. These NIR diagrams turn out to be
  comparable to the HST optical ones, both in depth and in photometric
  accuracy. The photometric errors for each bin of $K_s$ and $J$
  magnitudes are shown on the right side of the panels.}
\label{fig2}
\end{figure}

In order to obtain a clear definition of the evolutionary sequences in
the CMD, we selected only very well measured stars by imposing a
selection in the sharpness parameter. In particular, we divided
our sample in 0.5 magnitude-wide bins and for each bin we derived the
median value of the sharpness by applying an iterative
2$\sigma-$rejection (an example is shown in Figure \ref{sharp}). All
the stars satisfying this selection criterion and lying within 6$\sigma$ from
the median are shown in the ($K_s$, $J-K_s$) and ($J, J-K_s$) CMDs
plotted in Figure \ref{fig2}. These are the deepest and
highest-quality CMDs in the NIR ever obtained from the ground for this
cluster, but quite comparable with those obtained for NGC 6121 with HAWK-I@VLT (Very Large Telescope; \citealp{Lib14}) and for M15 with PISCES@LBT (Large Binocular Telescope; \citealp{Mon15}).

The NIR GEMINI CMDs presented in this study are fully
comparable, both in depth and resolution, to the optical ones obtained
by using different instruments on board HST, such as the WFPC2
\citep[][]{Hea00,Dal14} and the ACS/HRC \citep[][]{Sie11,Dal14}.  In
Figure \ref{fig2} the red HB at
$K_s$ $\sim$ 13.3 and a well defined RGB bump at about $K_s \sim
13.55$ are clearly distinguishable. Unfortunately, the stars lying along the brightest portion of
the RGB ($K_s< 14.0$) are saturated in all the available images. The
MS-TO region is well defined at $K_s\sim 17.5$ and the MS nicely extends
for more than 4 magnitudes down to $K_s \sim$ 21.5.  At odds with
\citet{Sie11}, we find no evidence of a broad sequence located below
the cluster MS at bluer colors, which has been interpreted as a
background feature, due to a tidal stream of the Sagittarius dwarf
galaxy, located behind the cluster along its line of
sight. The reason for such a discrepancy is likely ascribable to the
relatively small FOV of GSAOI compared to that of the ACS.  At $K_s
\sim$ 20 we clearly identify the so-called MS ``knee'' (MS-K), a
particular feature commonly defined as the reddest MS point in NIR
CMDs. This feature is due to the absorption of molecular hydrogen
induced by collisions \citep{Bon10} and flags the portion of the MS populated by
very low-mass stars (with $M<0.55 M_\odot$).  Until now, the MS-K has
been observed only by using HST optical+NIR images \citep{Mil12b, Mil14}
and a proper combination of HST and MCAO data (MAD@VLT - \citealp{Mor09}; PISCES@LBT - \citealp{Mon15} and GeMS - \citealp{Mas15,Tur15}).  
The detection shown in Figure \ref{fig2} 
is the first determination of the MS-K feature in a purely
NIR CMD obtained from the ground. The detailed study of this
feature and its use as a possible age indicator
\citep{Mas15,DiC15,Bon10} will be discussed in a forthcoming paper
(S. Saracino et al. 2016, in preparation).

\section{Absolute age determination}
\label{age}
In this Section we take advantage of the exquisite quality of the CMDs
shown in Figure \ref{fig2} to estimate the age of NGC 6624 by
using the luminosity of the MS-TO as an age indicator.

\subsection {Isochrone fitting method}
\label{age1}
In order to derive the age of the cluster via the isochrone fitting
method, three main cluster parameters are needed, namely the
metallicity, the distance and the reddening. 

\noindent{\it Metallicity -}  The most recent metallicity estimate of NGC 6624 is [Fe/H]$= -0.42 \pm 0.07$. This value has been derived indirectly from the new metallicity scale of \citet{Car09,Car10}, based on optical high-resolution spectra of about 20 GCs. 
Other estimates in the literature provide slightly lower metallicities:  
[Fe/H]$=-0.63$ (\citealp{CG97}, hereafter CG97;
\citealp{Hea00}). Recent high-resolution NIR spectroscopy by \citet{Val11} provides 
[Fe/H]$=-0.69 \pm 0.02$ and $[\alpha/Fe]\sim+0.39$. 

\noindent{\it Reddening - } $E(B-V)$ color excess estimates in the cluster
direction range from 0.25 to 0.33 \citep{Arm89,Hea00} with a fair agreement on $E(B-V)$ = 0.28 (\citealp{Va04a,Har96}, 2010 edition).

\noindent{\it Distance modulus - } Different estimates of the distance
modulus of NGC 6624 have been obtained: $(m-M)_0$ = 14.49 (\citealp{Har96}, 2010 edition), 
$(m-M)_0$ =14.40 \citep{Hea00}, and $(m-M)_0$ =14.63 \citep{Va04a}.

Both the reddening and the distance modulus were mainly derived in a
differential way, by comparing the CMD of NGC 6624 with those of other
clusters having the same metallicity and known values of $E(B-V)$ and
$(m-M)_0$. 

\begin{figure}
\epsscale{1.}
\plotone{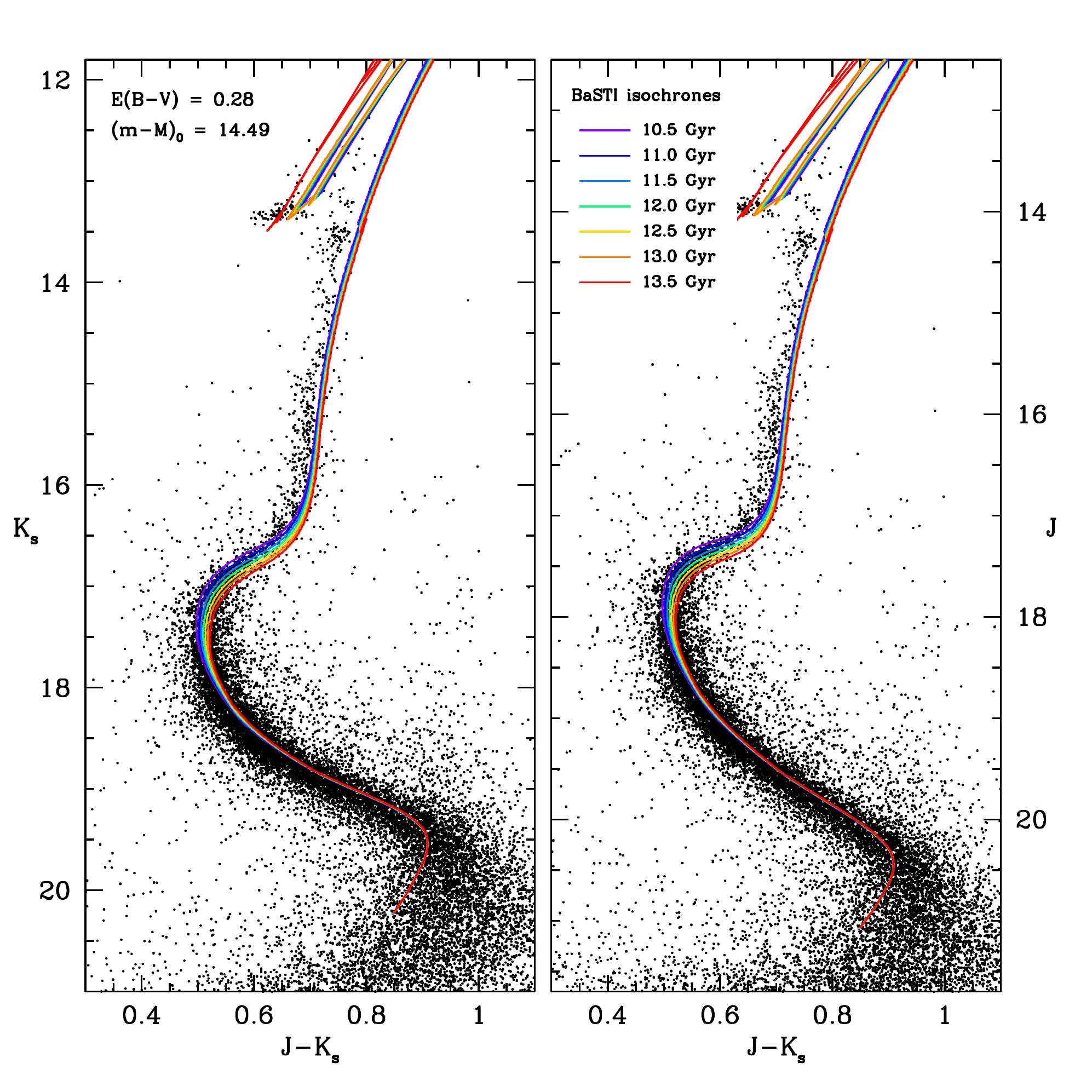}
\caption{($K_s, J-K_s$) and ($J, J-K_s$) CMDs of NGC 6624 CMDs with 
  overplotted a set of \texttt{BaSTI} \citep{Pie04} isochrones with ages ranging from
  10.5 Gyr up to 13.5 Gyr, in steps of 0.5 Gyr (see labels). }
\label{Basti1}
\end{figure}

In order to compare theoretical models with the NIR photometry of the cluster, we selected three different sets of $\alpha$-enhanced isochrones:
\begin{itemize}
\item[1.] A Bag of Stellar Tracks and Isochrones (\texttt{BaSTI}; \citealp{Pie04}) isochrones with $[\alpha/Fe]=+0.4$, Y = 0.259 and a mass-loss parameter $\eta$ = 0.4 \citep{Rei75}.
\item[2.] Dartmouth Stellar Evolutionary Database (\texttt{DSED}; \citealp{Dot07}) isochrones with $[\alpha/Fe]=+0.4$ and Y = 0.2583.
\item[3.] Victoria-Regina (\texttt{VR}; \citealp{Van14}) isochrones with $[\alpha/Fe]=+0.4$ and Y = 0.2583.
\end{itemize}
For each model we adopted [Fe/H] = -0.60, a distance modulus of $(m-M)_0 = 14.49$ ($d=7.9$ Kpc) and a
color excess $E(B-V) = 0.28$\footnote{The color excess in the NIR filters $E(J-K_{s})$ has been derived 
by adopting the extinction coefficients $A_{J}/E(B-V)=0.899$ and $A_{K_s}/E(B-V)=0.366$ from \citet{CV14}}. These  
parameters well reproduce
the main evolutionary features in the high-quality optical HST CMD by \citet{Dal14}. 
However, the impact of adopting different values for
these parameters will be discussed at the end of this Section. \\\\ 
\texttt{BaSTI} {\it isochrones -}
Figure \ref{Basti1} shows a set of seven isochrones with appropriate
 different ages ranging from 10.5 to 13.5 Gyr, stepped by 0.5 Gyr, overplotted to the
data of NGC 6624 in the ($K_s, J-K_s$) and ($J, J-K_s$) CMDs, by
assuming the quoted distance and reddening. The \texttt{BaSTI} NIR colors and magnitudes are on the Johnson-Cousins-Glass photometric system, so they were converted first on the \citet{BB88} system and then, by using the transformations of \citet{Car01}, into the 2MASS photometric system \citep{Cut03}.\\  As can be seen, the
agreement between \texttt{BaSTI} models 
and our data is quite good in the extended MS and in the TO region, while it is not completely satisfactory
in terms of color at the level of the MS-K (at $K_s >
19.5$)\footnote{However, it should be noted that the \texttt{BaSTI} models do
  not sample the entire extension of the MS since they are truncated
  at $M = 0.5 M_\odot$.} and in the lower part of the RGB. 
A mismatch between models and data appears also at the brightest portion of the RGB, at $K_s< 14$ but in this case it is likely due to non-linearity and
saturation problems that mainly affect the $K_s$ band photometry.
In spite of this, both the luminosity of the RGB
bump and the ZAHB level are well reproduced (see Figure \ref{Basti1}).

\begin{figure}
\epsscale{1.}
\plotone{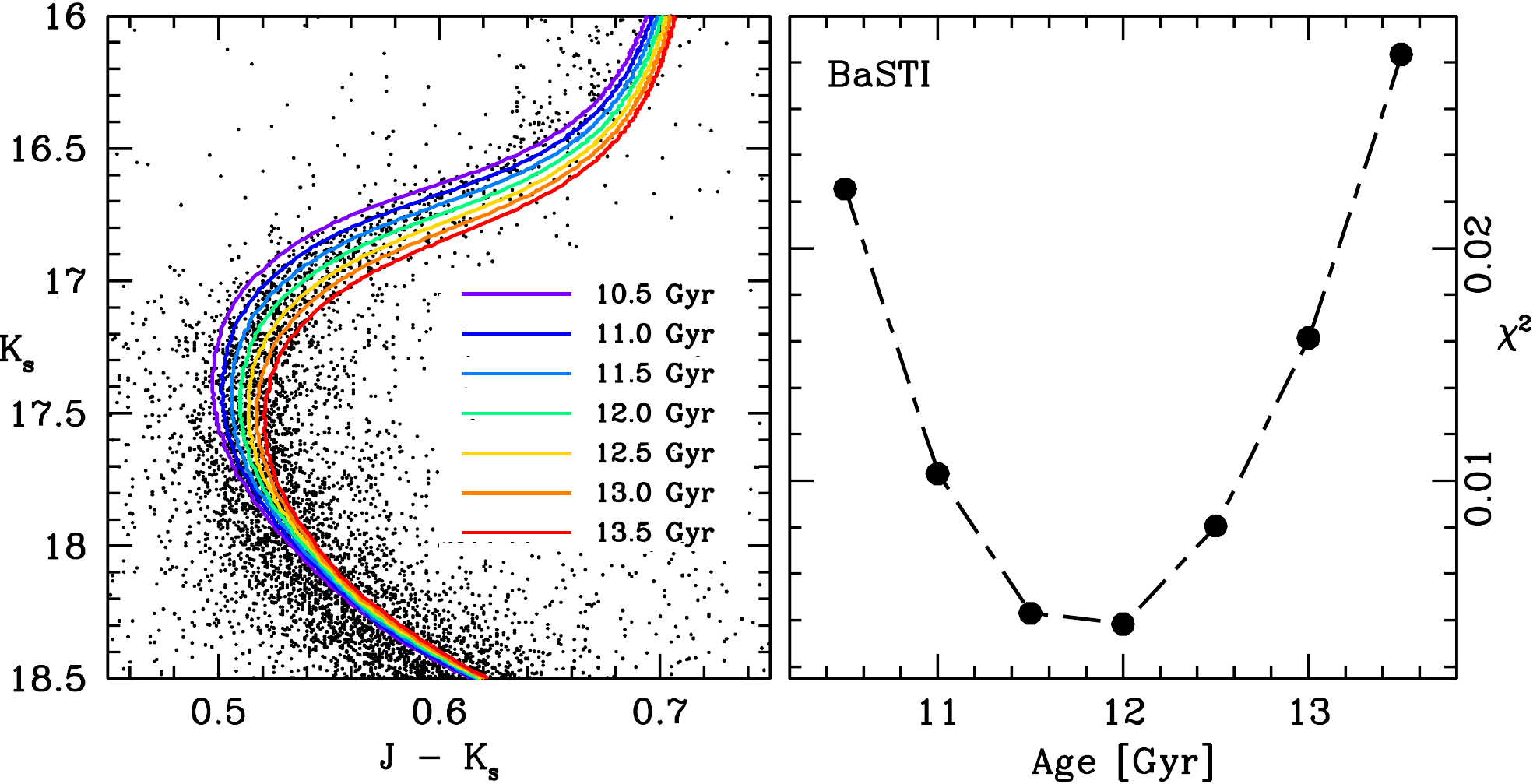}
\caption{{\it Left panel:} zoomed MS-TO region of the ($K_s,
  J-K_s$) CMD, with the selected set of \texttt{BaSTI} isochrones (with
  different ages) overplotted. {\it Right panel:}  $\chi^2$
  parameter as a function of isochrone ages
  considered in the {\it Left Panel}. A well defined minimum
  identifies the best-fit isochrone (with $t_{age}$ = 12.0 Gyr).  }
\label{TO}
\end{figure}

\begin{figure}
\epsscale{1.}
\plotone{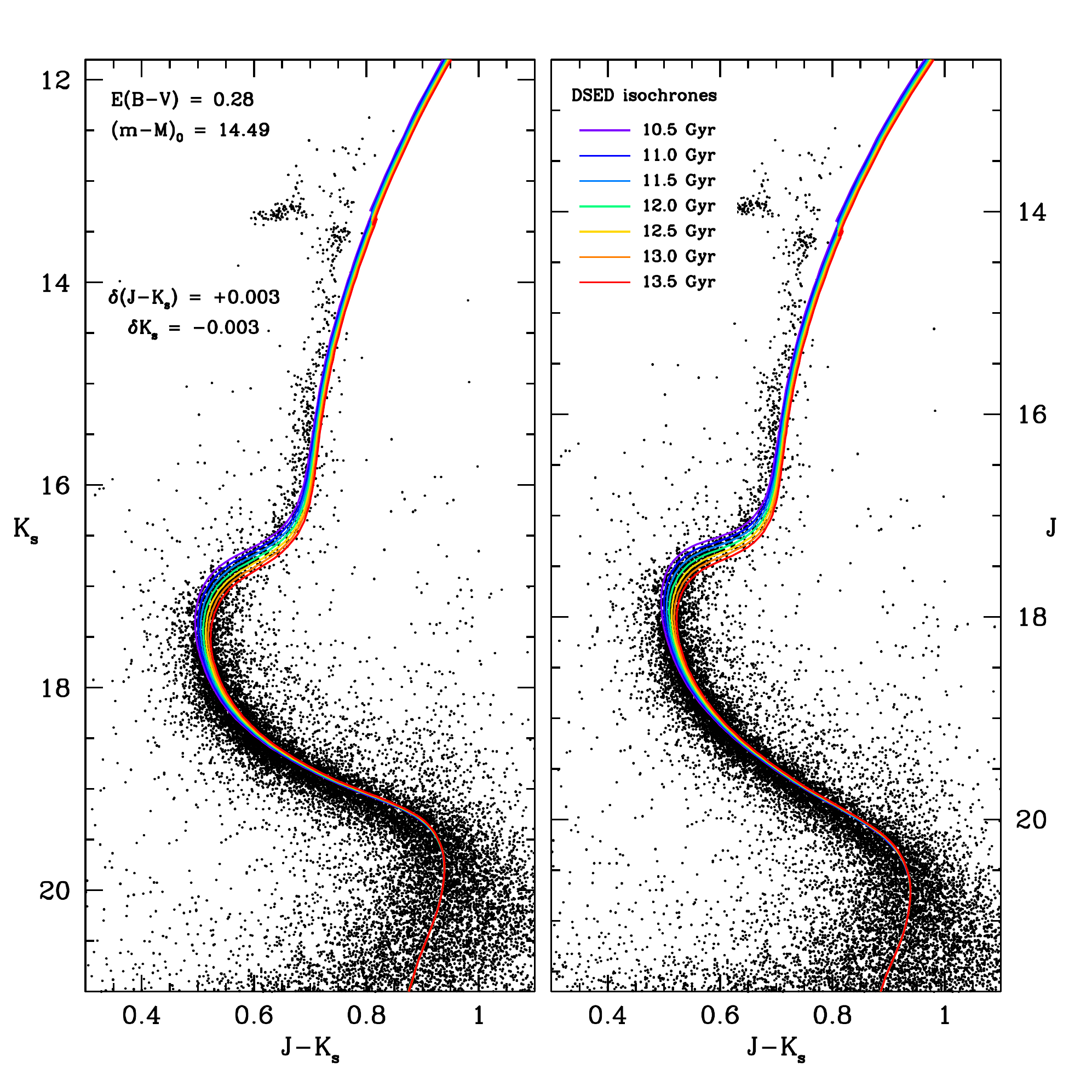}
\caption{The same as in Figure \ref{Basti1} but for \texttt{DSED} isochrones \citep{Dot07}.}
\label{Dotter}
\end{figure}

\begin{figure}
\epsscale{1.}
\plotone{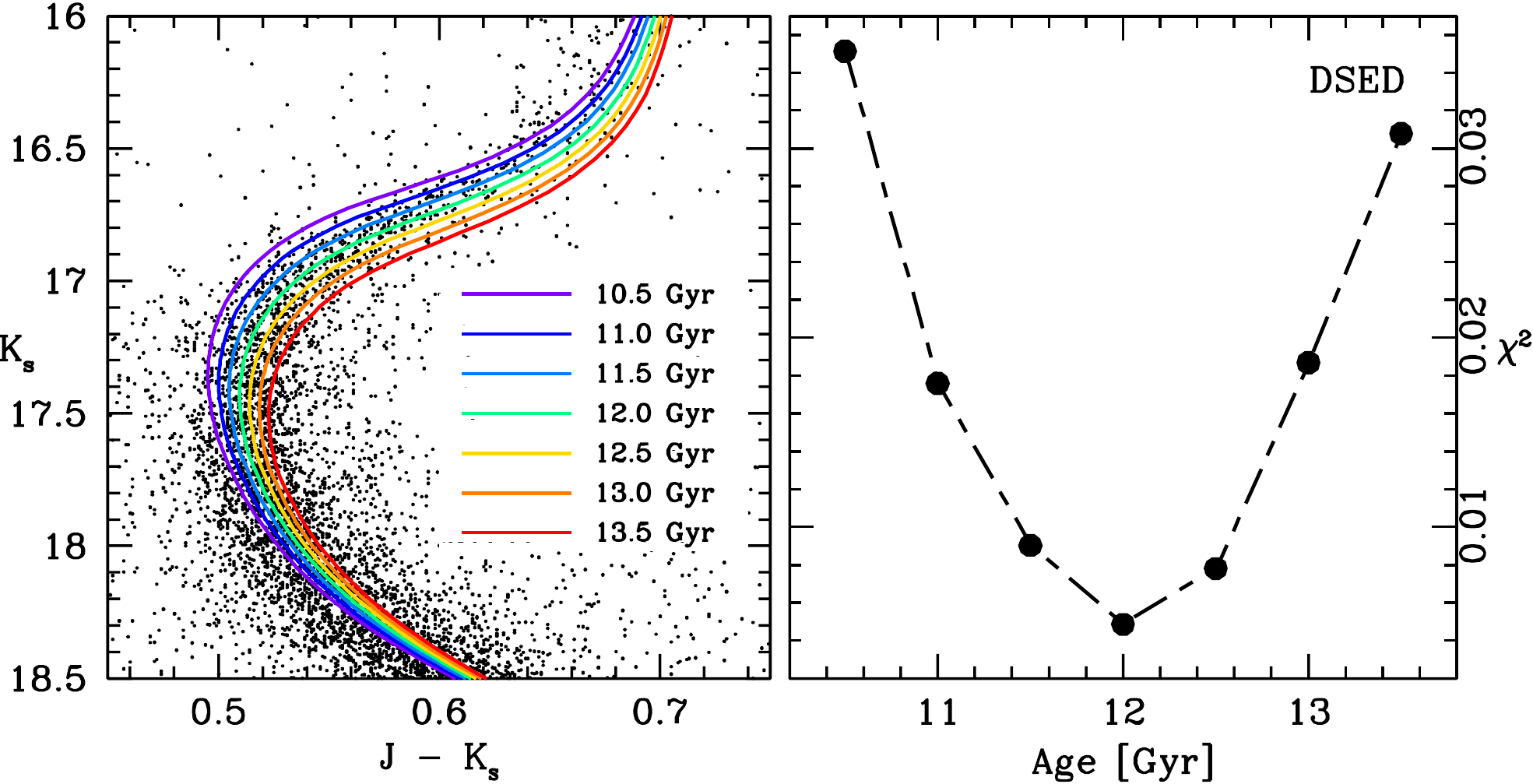}
\caption{The same as in Figure \ref{TO} but for \texttt{DSED} isochrones.}
\label{DSEDchi}
\end{figure}

\begin{figure}
\epsscale{1.}
\plotone{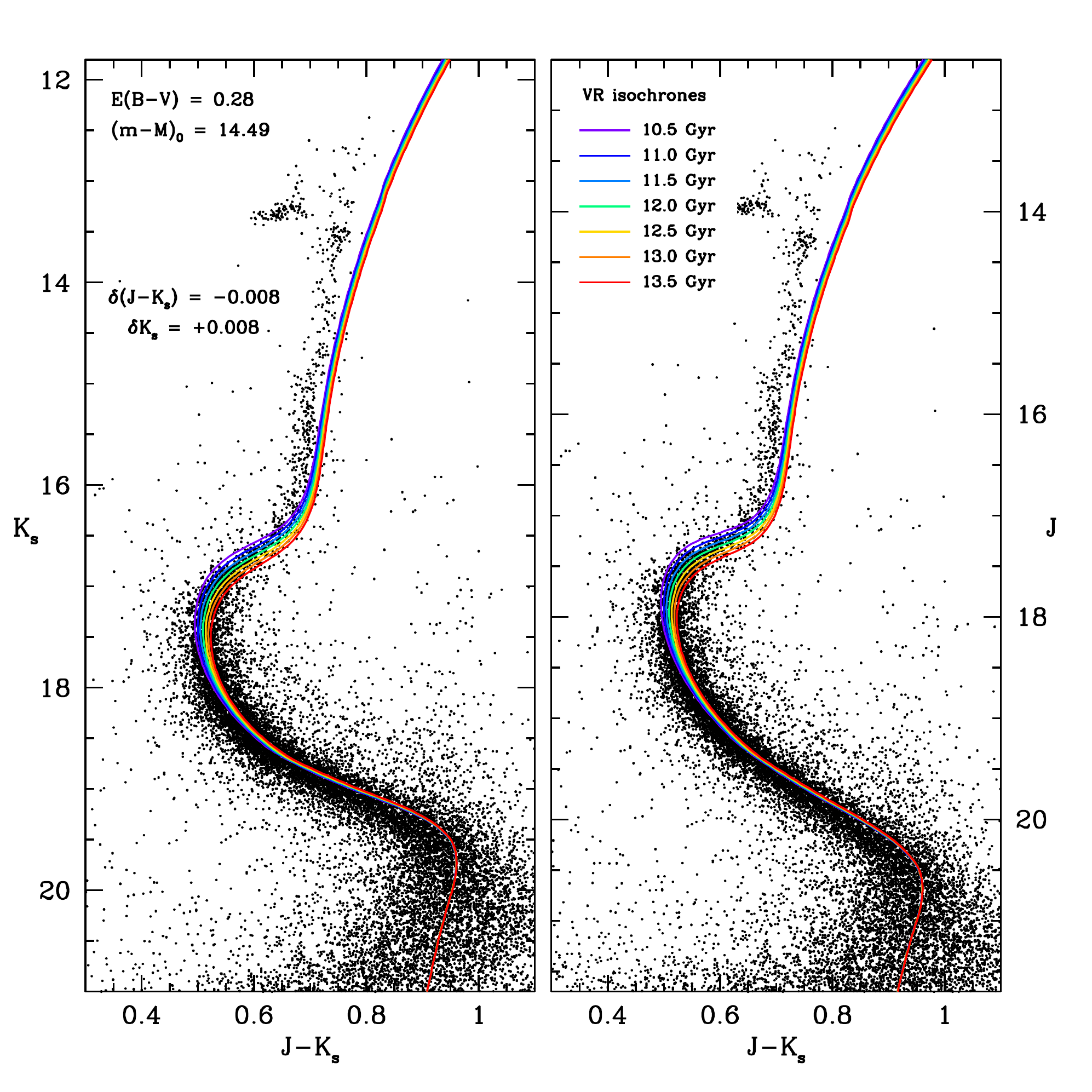}
\caption{The same as in Figure \ref{Basti1} but for \texttt{VR} isochrones \citep{Van14}.}
\label{viri}
\end{figure}

Focussing on the TO region, in order to identify the isochrone that
best reproduces the observations, we performed a $\chi^2$
analysis. The $\chi^2$ parameter has been computed by selecting a
subsample of stars in the MS-TO region, in the magnitude range $16.4 <
K_s < 17.8$, where the isochrone shape is particularly sensitive to
age variations. It has been defined as: $\chi^2 = \sum[(O_K-E_K)^2 + (O_{colJK} - E_{colJK})^2]$. We computed, iteratively, the minimum distance both in magnitude ($O_K$) and in color ($O_{colJK}$) between each star in our sample and the corresponding values $E_K$ and $E_{colJK}$ read along the isochrone. The
result is shown in the right panel of Figure \ref{TO}. As can be seen,
a well defined minimum is visible, indicating a best-fit isochrone
with an absolute age of $t_{age}=12.0\pm0.5$ Gyr.\\\\
\texttt{DSED} {\it isochrones -}
Following the same approach adopted for the \texttt{BaSTI} isochrones, Figure \ref{Dotter} shows the comparison between the observed cluster photometry and the \texttt{DSED} isochrones \citep{Dot07}. A color offset $\delta(J-K_s)=+0.003$ (corresponding to $\delta K_s=-0.003$) was needed to reconcile these evolutionary models to the data, in agreement with what found by \citet{Coh15}.
\texttt{DSED} isochrones do a good job of reproducing the observed CMD morphology of NGC 6624, from the SGB level down to the MS-K. The RGB instead shows the same discrepancies already seen for the \texttt{BaSTI} isochrones, with the only difference that in this case, even the magnitude of the RGB bump is not well reproduced. This is likely due to the fact that different models treat overshooting from convective cores in different ways. 
A $\chi^2$ test has been performed also for \texttt{DSED} models and the minimum value has been obtained for an absolute age of $t_{age}=12.0\pm0.5$ Gyr. This result is shown in Figure \ref{DSEDchi}, together with a zoom into the MS-TO/SGB regions of NGC 6624.\\\\
\texttt{VR} {\it isochrones -}
Finally, the ($K_s, J-K_s$) and ($J, J-K_s$) CMDs of NGC 6624 have been compared to the new \texttt{VR} isochrones (\citealp{Van14}) (see Figure \ref{viri}). According to the results of \citet{Van13} and \citet{Coh15} we needed to apply a color shift of -0.008 in order to have a good match between models and data.
The MS-TO and the SGB level are well reproduced, as well as the shape of the MS-K. At the base of the RGB instead, \texttt{VR} models are systematically too red compared with the observed photometry. By performing a $\chi^2$ test on the \texttt{VR} isochrones, we obtained a slightly older absolute age of $t_{age}=12.5\pm0.5$ Gyr, however in agreement, within the uncertainties of the method, with the ages derived from \texttt{BaSTI} and \texttt{DSED} models.

Figures \ref{Basti1}, \ref{Dotter} and \ref{viri} show that there is a mismatch between all models and observation starting from the RGB base. As discussed in detail by \citet{Sal07}, \citet{Bra10} and \citet{Coh15}, this is a well known and long-standing problem. It may be related to some issues depending on the $T_{eff}$ - color transformations in the IR, and possibly caused by uncertainties in the model atmospheres, like treatment of absorption lines as a function of gravity and to the abundance of some specific elements.\\ The impact of the adopted chemical composition on the overall quality of the fit has been tested. In particular, we compared isochrones with [Fe/H] = -0.42 and [$\alpha$/Fe] = +0.2 adopted by \citet{Van13} for their age derivation of NGC 6624 and [Fe/H] = -0.60 \& [$\alpha$/Fe] = +0.4 that we have used for our age derivation.
For a fixed age, the two isochrones ([Fe/H] = -0.60 \& [$\alpha$/Fe] = +0.4 and [Fe/H] = -0.42 \& [$\alpha$/Fe] = +0.2) exactly match, showing only a slight difference at the MS-K level. We have also considered the effect of the abundance variations. \citet{Coh15} showed that a modest helium enhancement of Y=0.04 is not able to solve the observed disagreement between models and data at the SGB/RGB level. We verified that the overall match still remains unsatisfactory even adopting isochrones with an helium content as large as Y=0.35. In particular, while the increase of helium results in a better fit at the RGB base, it does not allow to properly reproduce the MS. 
Finally, we tried to test the impact of adopting different values of reddening and distance on the determination of the age of the cluster. With this aim we performed the same $\chi^2$ analysis as before by assuming the values of \citet{Hea00} and \citet{Va04a}. This analysis gives age values of $11.0\pm0.5$ Gyr and $10.5\pm0.5$ Gyr respectively. However a visual check reveals that isochrones having such extreme values of E(B-V) and/or $(m-M)_0$ do not reproduce properly the main evolutionary sequences (MS-TO/SGB level, HB level and RGB bump) of the cluster. Hereafter, for our analysis we will adopt the absolute age of 12 Gyr derived from \texttt{BaSTI}, the model which better reproduce the overall shape of the CMD of NGC 6624.

\subsection{Comparison with previous results}
The age of NGC 6624 has been widely debated in recent years.  As
shown in Figure \ref{age_conf}, the value determined in this work is
in good agreement, within the errors, with most of the previous
estimates quoted in the literature.

The largest discrepancy is found with respect to the value determined
by \citet{Hea00}. By comparing the CMD of NGC 6624 with those of NGC
6637 and 47 Tucanae (having a similar metallicity), the authors
established that these clusters have the same age ($14 \pm 1$ Gyr) 
by using the Yale isochrones.  A relatively old age ($13.00 \pm 0.75$
Gyr) was also found by \citet{Dot10}, analyzing \texttt{DSED} models with
the following chemical composition: [Fe/H]$= -0.50$ (by assuming the
scale by \citealp{ZW84}, hereafter ZW84) and [$\alpha$/Fe]$=0$.  An
absolute age of $10.6 \pm 1.4$ Gyr, in quite good agreement with the
one estimated in this work (within their large error bars), was instead obtained by \citet{SW02}, both
for a metallicity of [Fe/H]$= -0.50$ (ZW84 scale) and for
[Fe/H]$=-0.70$ (CG97 scale).  \citet{MW06} derived for NGC 6624 two
different ages: 12 Gyr from the model best reproducing the MS-TO
region (as in our case), and a lower limit of 10.5 Gyr, from the isochrone providing an
acceptable fit to all the investigated age indicators.
From the isochrone fitting and the ZAHB loci, \citet{Van13} derived
the ages of 55 GCs for which the optical HST/ACS photometry was
publicly available. In the case of NGC 6624 the absolute age resulted
to be 11.25 $\pm$ 0.5 Gyr, in good agreement with our value. Finally the most recent study of the
cluster was performed by \citet{Roe14}, within a detailed review on
the state of the art about age and metallicity determinations for a
sample of 41 Galactic GCs. The authors quote an age of 12.5 $\pm$ 0.9
Gyr, which is again consistent with the value obtained in this
work.

\section{The Luminosity and Mass Function of NGC 6624}
\label{luminosity}
We have studied the LF and the MF of MS stars in NGC 6624 and their radial variations within the FOV of
GSAOI. This kind of studies has been probed to be an efficient tool to study the effect of cluster internal dynamics on stars in a wide range of masses, including the faint-end of the MS where most of the cluster mass lies. In relaxed systems, the slopes of the MF and LF are expected to vary as a function of the distance from the cluster center, with indexes decreasing as distance increases, because of the different effect of mass segregation. Moreover, the radial variation in the stellar MF of star clusters allow to derive crucial information about the dynamical history of the system (including the amount of mass loss suffered by the clusters, e.g. \citealp{deM07}, \citealp{VH97}) and variation in the stellar initial mass function as recently shown by \citet{WV16}. In this context, NGC 6624 is a quite interesting case because it has been classified as a post-core collapse cluster
\citep{Tra95}, hence it has already experienced some of the most
energetic and advanced phenomena known to occur during the dynamical
evolution of dense stellar systems \citep{Mey97}.

\subsection{Artificial Star Experiment}
\label{artificial}
In order to derive a meaningful MS-LF, it is important to take into
account a number of effects (such as blends, photometric errors
and stellar crowding) that can limit the photometric accuracy,
especially when deriving complete samples of faint stars. In order to
assess the impact of these effects, we performed a set of artificial
star experiments following the prescriptions reported in
\citet{Bel02b} and \citet{Dal11,Dal15}. The main steps of this
procedure are briefly summarized below. We initially defined a mean
ridge-line (MRL) in the ($K_s, J-K_s$) plane by considering 0.5
mag-wide bins in $K_s$ and by selecting the corresponding median value
in color after a 2$\sigma-$clipping rejection (see Table 1 and the red line in Figure \ref{sim_conf}, left panel). Then, we generated a
catalog of simulated stars with a $K_s$-input magnitude ($K_{s, {\rm
    in}}$) extracted from a LF modeled to reproduce the observed LF.
To each star extracted from the LF, we finally assigned a $J_{in}$
magnitude by means of an interpolation along the MRL of the
cluster. In order to avoid artificially increasing the crowding
conditions, only one artificial star was simulated in each run within
a $20\times20$ pixels cell (more than 5 times the typical FWHM of stars
on the images).  In addition,
we imposed each simulated star to have a minimum distance of about 50
pixels from the edges of each chip. By imposing these criteria and
adopting the coordinate transformations discussed in Section
\ref{obs}, artificial stars were added to the 13 $J$ and 14 $K_s$ images
 by using the \texttt{DAOPHOT II/ADDSTAR} software. This procedure
was performed individually on each chip and the artificial stars thus
obtained were analyzed by using the same PSF models and the same
reduction process (including the same selection criteria) adopted for
the real images, as fully described in Section \ref{obs}. A total of
$\sim$ 48000 artificial stars per chip have been simulated. Figure
\ref{sim_conf} shows a comparison between the simulated ({\it left
  panel}) and the observed CMD ({\it right panel}). It is quite
evident that the two CMDs are fully compatible over the entire
magnitude range covered by the observations, in particular the
simulated MS (for $K_s> 16$) shows the same spread in color as the
observed one, thus confirming that the procedure is fully appropriate.

The artificial star catalog thus obtained was used to derive the
completeness curves as the ratio ($\Gamma=N_{\rm out}/N_{\rm in}$)
between the number of stars recovered after the photometric reduction
($N_{\rm out}$) and the number of simulated stars ($N_{\rm in}$) in
each magnitude bin. The completeness curves as a function of the $K_s$
magnitudes for three different radial bins ($0\arcsec \le r <
15\arcsec$, $15\arcsec \le r < 30\arcsec$ and $r \ge
30\arcsec$, where $r$ is the distance from the cluster
center\footnote{We recomputed the cluster center $C_{\rm grav}$ of NGC
  6624 by applying the algorithm of \citet{Cas85} to evaluate, through
  an iterative procedure, the ``density center'' of the stellar
  positions, i.e. their average position weighted by the local number
  density (for more details, see \citealt{Lan07,Lan10}).  Using
  various initial searching radii, ranging from $5\arcsec$ to
  $20\arcsec$, and applying a cut in magnitude $K_s$ = 18 to avoid
  incompleteness, we finally found $\alpha_{J2000}=18^h 23^m 40^s
  .52$, $\delta_{J2000}= -30^\circ 21\arcmin 40\arcsec .25$ with an
  uncertainty of about $0.2\arcsec$.}) are shown in Figure \ref{compl}. This clearly testify
to a high (90-100\%) photometric completeness of the
GEMINI catalog down to $K_s \sim 19-20$, depending on the distance from the center.

\subsection{Luminosity and Mass functions}
\label{lum}
In order to derive the MS-LF of NGC 6624 we adopted the radial bins
defined for the determination of the completeness curves (see Section \ref{artificial}). 
Starting from the ($K_s, J-K_s$) CMD, we selected a sample
of bona-fide MS stars defined as those stars located within
2.5$\sigma$ from the MRL, where $\sigma$ is the combined photometric
uncertainty in the $K_s$ and $J$ bands. For each radial interval, we
considered only the stars with $K_s$ magnitude ranging between 16 and
the value corresponding to a completeness factor $\Gamma\sim 0.5$,
namely $K_s$ = 20.35 for $0\arcsec \le r < 15\arcsec$,
$K_s=20.45$ for $15\arcsec \le r < 30\arcsec$ and
$K_s=20.55$ for $r \ge 30\arcsec$.  
The observed LF obtained in each
radial interval was thus corrected for incompleteness by adopting the
appropriate value of $\Gamma$ in each bin of magnitude. The
completeness-corrected LF was then decontaminated from the effect of
field stars by using the Besan\c{c}on simulation of our Galaxy
\citep{Rob03} for a region of about 2.5\arcmin$\,$ squared centered on
NGC 6624, rescaled on the GSAOI FOV. For each magnitude bin, the density of field stars has
been estimated and then subtracted (see for example
\citealt{Bel99}). In Figure \ref{LF} the completeness-corrected and
background-subtracted LFs of NGC 6624 in different radial bins are
shown. In order to compare the LFs obtained at different distance from
the cluster center, the LFs have been normalized to the one obtained
for the innermost region in the magnitude range $16.0 < K_s <
17.5$. The comparison shows that the LFs are clearly different, with
the LF in the outer regions of the cluster showing
an overabundance of low-luminosity stars with respect to those 
in the innermost radial bin, consistent with the effect of mass segregation. This is in agreement with results by \citet{Gol13}.

Following a similar approach, we have also derived the MF of NGC
6624. Masses have been estimated from the \texttt{BaSTI} isochrone that
best-fits the CMD, as discussed in Section \ref{age1}. This provides a
MS-TO mass of $0.88 M_\odot$. The MF covers a range in mass from $0.9
M_\odot$ down to lower limits corresponding to $\Gamma\approx 0.5$: $M
= 0.49 M_\odot$ for $0\arcsec \le r < 15\arcsec$, $0.47 M_\odot$
for $15\arcsec \le r < 30\arcsec$ and $0.45 M_\odot$ for $r \ge
30\arcsec$.  The MFs are shown in Figure \ref{MF}, with the same color
code as in Figure \ref{LF} to indicate different distances from the
cluster center. We used the MF of the innermost radial
bin as a reference and we normalized the more external ones by using
the number counts in the mass range $0.8 \le M/M_{\odot} < 0.9$. As
appear in Figure \ref{MF} and in agreement with what we obtained from
the analysis of the LFs, the MFs show a clear variation of their slopes, 
flattening moving outwards. 
Such a trend clearly demonstrates that this cluster has already experienced a significant degree of mass segregation. In general, all GCs are expected to show evidence of mass segregation, because of their short relaxation timescales \citep[e.g.][]{Pau10}. However, recent results have shown that there are some exceptions: see the case of $\omega$-Centauri (\citealp{Fer06a}), NGC 2419 (\citealp{Dal08}, \citealp{Bel12}), Terzan 8 and Arp 2 (\citealp{Sal12}), Palomar 14 (\citealp{Bec11}) and NGC 6101 (\citealp{Dal15}). These examples suggest that either some mechanisms able to suppress mass segregation may occur in some GCs or that theoretical relaxation times may suffer of significant uncertainties. In this context, the study of the dynamical state of a cluster becomes quite interesting.

\section{Summary \& Conclusions}
\label{concl}
This work is focused on NGC 6624, a metal-rich globular cluster
located in the Galactic bulge. By combining the exceptional
capabilities of the adaptive optics system GeMS with the high
resolution camera GSAOI on the GEMINI South telescope, we obtained the
deepest and most accurate NIR ($K_s, J-K_s$) and ($J, J-K_s$) CMDs
ever obtained from the ground for NGC 6624. The quality of the photometry turns out
to be fully competitive with the optical photometry from the HST. 
The derived CMDs span a range of more than 8 magnitudes,
allowing to identify all the well known evolutionary sequences, from
the HB level to the MS-TO point down to below the MS-K (detected at $K_s$
$\sim $ 20), a feature observed so far only rarely in the optical band
and identified for the first time here in a purely infrared CMD. We
took advantage of our high-resolution photometry to get an accurate
estimate of the absolute age of NGC 6624, which is still quite debated
in the literature. By adopting the MS-TO fitting method, we determined
an absolute age of about 12.0 $\pm$ 0.5 Gyr for the cluster.

Taking advantage of this high-quality sample, we studied the MS-LF and
MF at different distances from the cluster center. The level of
completeness of the MS sample has been evaluated from artificial star
experiments and turns out to be larger than 50\% down
to $K_s\sim 20.3$ at any distance from the cluster center. The
completeness-corrected and field-decontaminated LFs and MFs show
significant signatures of mass segregation.  In fact, moving from the
innermost region of the cluster to the outskirts, the number of
low-mass stars gradually increases compared to high-mass stars. This
result confirms that NGC~6624 is a dynamically old cluster, already
relaxed.

The data obtained for NGC 6624 clearly show that, under favorable
conditions (for example the seeing of the observing night or the NGSs
magnitude), the GeMS/GSAOI system is able to provide images with similar
spatial resolution and photometric quality
as HST in the optical bands. 

\begin{appendices}
\section{Testing stellar variability with the GeMS/GSAOI system}
We could also test the stability and accuracy of the GeMS + GSAOI
system in performing stellar variability studies. To do this, we
focused on the core of NGC 6624. In this region \citet{Deu99}
discovered an exotic object (Star1) which has been classified as a
quiescent cataclysmic variable or a LMXB. The optical counterpart to
this object, named COM\_Star1, has been identified by \citet{Dal14} to
be a star showing a clear sinusoidal light modulation and an orbital
period of $P_{orb}$ $\approx$ 98 minutes. We identified COM\_Star1
in our GEMINI $J$ and $K_{s}$ images (see Figure \ref{ComStar1}) by
using the coordinates reported by \citet{Dal14}. It appears as a relatively bright star
at $K_s$ $\sim$ 17.76 and $(J-K_s) \sim 0.69$ outside the MS on the red
side (it is marked with a red circle in the top panel of Figure
\ref{cmdComStar1}). Since the object is visible in all GEMINI images
(13 in $J$ and 14 in $K_{s}$), we could investigate its variability in
the infrared bands.  Its light curve (red dots in the {\it bottom
  panel} of Figure \ref{cmdComStar1}) shows a luminosity variation
with an amplitude of $\sim$ 0.2 mag and is well folded with the same
period ($P_{orb}=98$ min) found by \citet{Dal14}. For sake of comparison, the light curve of a
genuine MS star (StarB) with comparable luminosity (blue circle in
Figure \ref{StarB}) is shown in Figure \ref{cmdComStar1}, {\it bottom
  panel}. As expected, this star is a ``normal" non-variable MS, in
fact its light curve does not show any evidence of flux modulation
($\sigma_K=0.017$ mag).\\
The constancy of the StarB magnitude despite variation of seeing,
airmass and PSF shape in the images, implies that the system is
largely stable and it can be used to successfully reveal also very
small flux variations (of the order of a few 0.01 mag). This test
allow us to conclude that the GeMS/GSAOI system indeed has such a high performance 
and can be successfully used also for stellar variability
studies.
\end{appendices}

\clearpage
\acknowledgments  We thank the anonymous referee for the careful reading of the paper, and for the useful comments/suggestions. This research is part of the project {\it Cosmic-Lab}
(http://www.cosmic-lab.eu) funded by the European Research Council
under contract ERC-2010-AdG-267675. S.S. acknowledges the {\it ``Marco
  Polo Project"} of the Bologna University for grant support and the
University of Concepci\'on for the warm hospitality during her stay
when part of this work has been carried out.
D.G. and S.V. gratefully acknowledge support from the Chilean 
BASAL Centro de Excelencia en Astrof\'isica
y Tecnolog\'ias Afines (CATA) grant PFB-06/2007. F.M. gratefully
acknowledges the support provided by Fondecyt for project
3140177.
L.O. acknowledges the PRIN-INAF 2014 CRA 1.05.01.94.11: 
``Probing the internal dynamics of globular clusters. The first
comprehensive radial mapping of individual star kinematics with the 
new generation of multi-object spectrographs'' (PI: L.
Origlia). R.E.C. acknowledges funding from Gemini-CONICYT Project 32140007. C.M.B. acknowledges support from project Fondecyt Regular 1150060. S.S. and E.D. thank Maurizio Salaris for useful discussions and inputs about the stellar evolutionary models.

\clearpage
%%%%%%%%%%%%%%%%%%%%%%% tables %%%%%%%%%%%%%%%%%%%%%%%%%
\begin{table}[!ht]
\begin{center}
\caption{Mean Ridge-Line of NGC 6624 in the ($K_s, J-K_s)$ CMD.}
\label{tab1}
\footnotesize
\begin{tabular}{ll}
\\
\hline
\hline
$\;\;\;K_{s}$ & $J - K_s$ \\
(mag) & (mag)\\
\hline
12.413 & 0.742 \\
12.913 & 0.737 \\
13.413 & 0.727 \\
13.913 & 0.723 \\
14.413 & 0.714 \\
14.913 & 0.706 \\
15.413 & 0.696 \\
15.913 & 0.680 \\
16.413 & 0.644 \\
16.913 & 0.547 \\
17.413 & 0.519 \\
17.913 & 0.528 \\
18.413 & 0.588 \\
18.913 & 0.709 \\
19.413 & 0.897 \\
19.913 & 0.947 \\
20.413 & 0.937 \\
20.913 & 0.877 \\
21.413 & 0.817 \\
\hline
\end{tabular}
\end{center}
\end{table}

%%%%%%%%%%%%%%%%%%%%%%%% figures %%%%%%%%%%%%%%%%%%%%%%%%%%%%%%%%%%%

\clearpage
\begin{figure}
\epsscale{1.}
\plotone{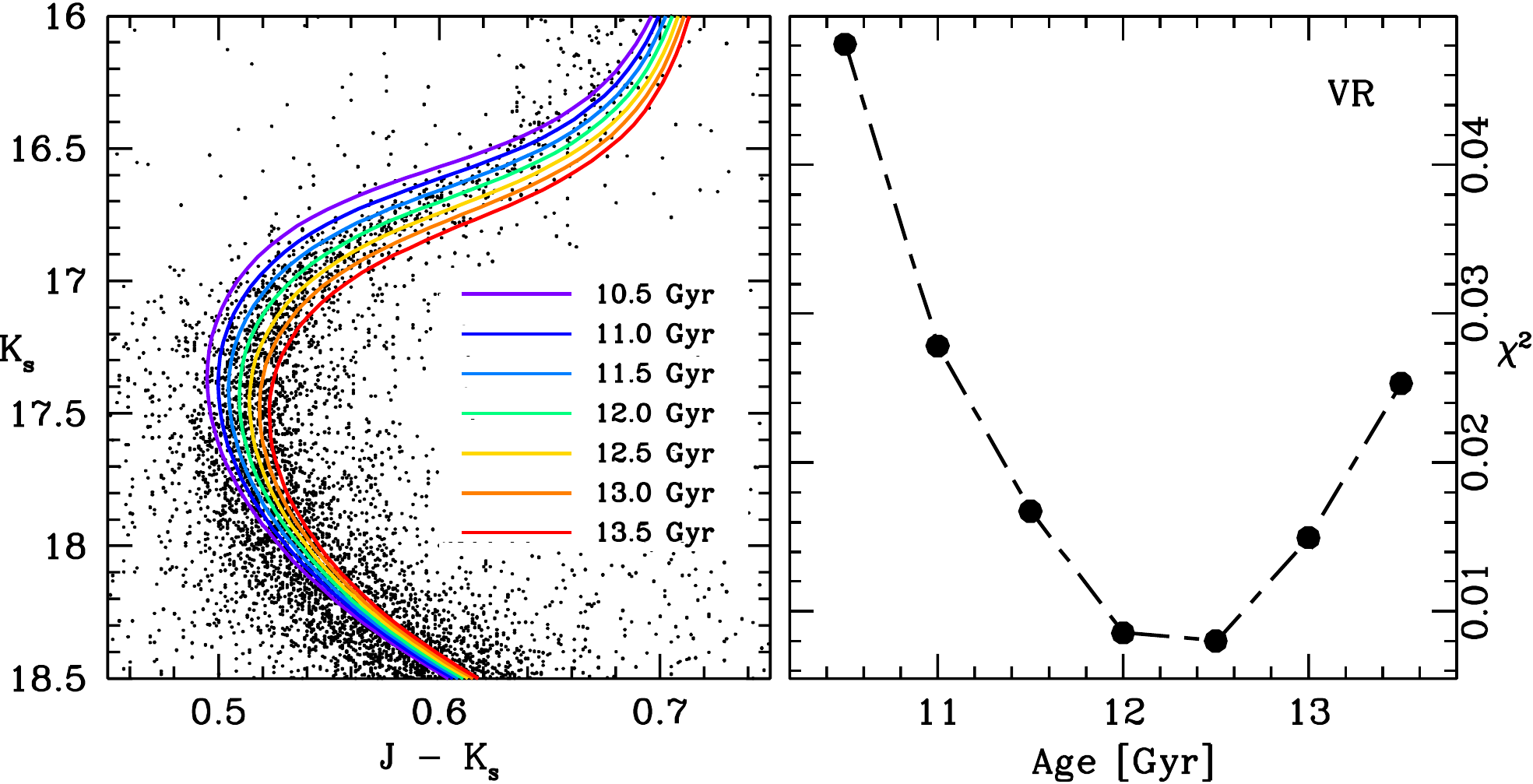}
\caption{The same as in Figure \ref{TO} but for \texttt{VR} isochrones.}
\label{VRchi}
\end{figure} 

\begin{figure}
\epsscale{1.}
\plotone{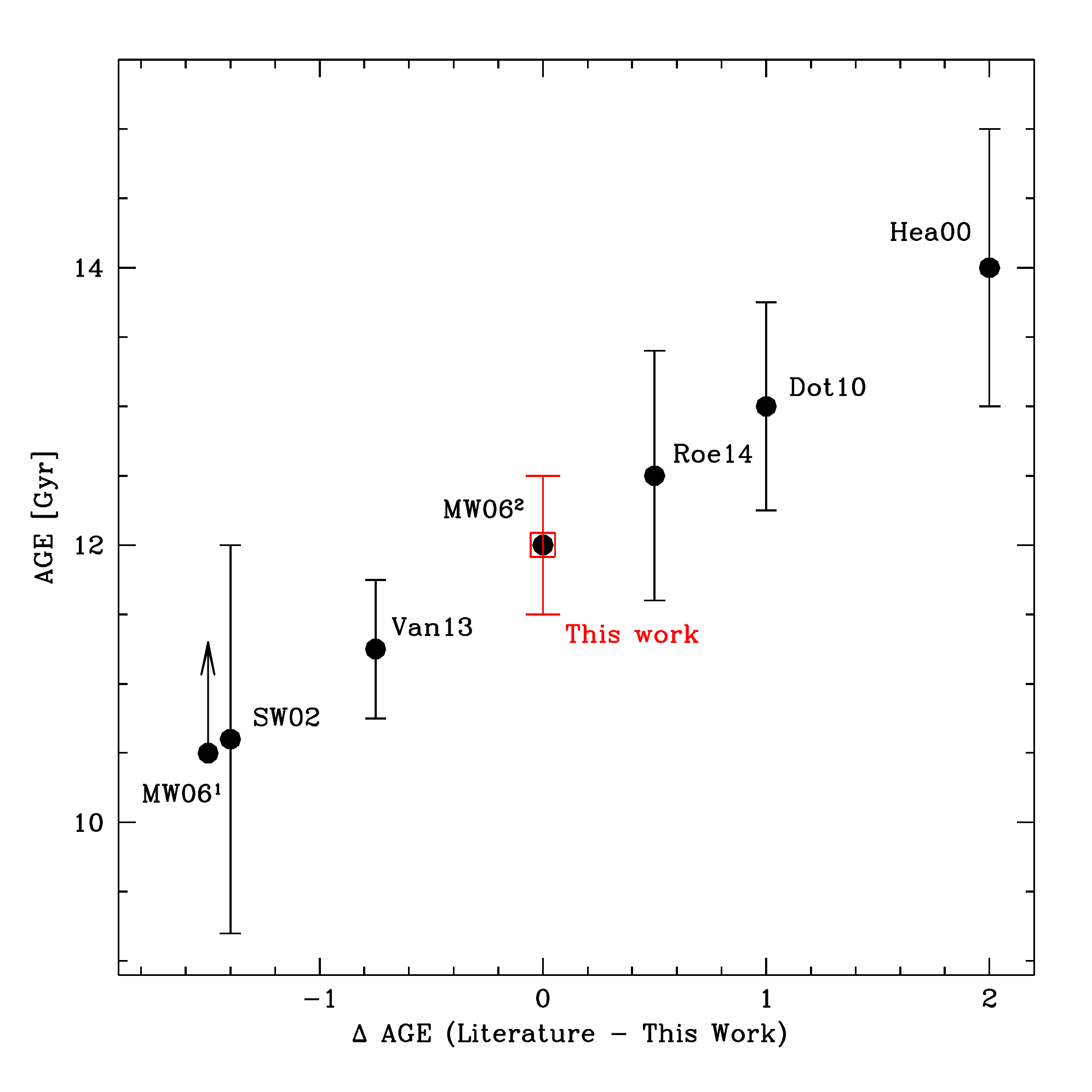}
\caption{Age estimates of NGC 6624 determined in previous
  studies, compared to that obtained in this work (red point). The
  acronyms shown in the figure are so defined: {\bf MW06$^{1}$} \& {\bf MW06$^{2}$}
  \citep{MW06}, {\bf SW02} \citep{SW02}, {\bf Van13} \citep {Van13},
        {\bf Roe14} \citep{Roe14}, {\bf Dot10} \citep{Dot10} and {\bf
          Hea00} \citep{Hea00}.}
\label{age_conf}
\end{figure} 

\begin{figure}
\epsscale{1.}
\plotone{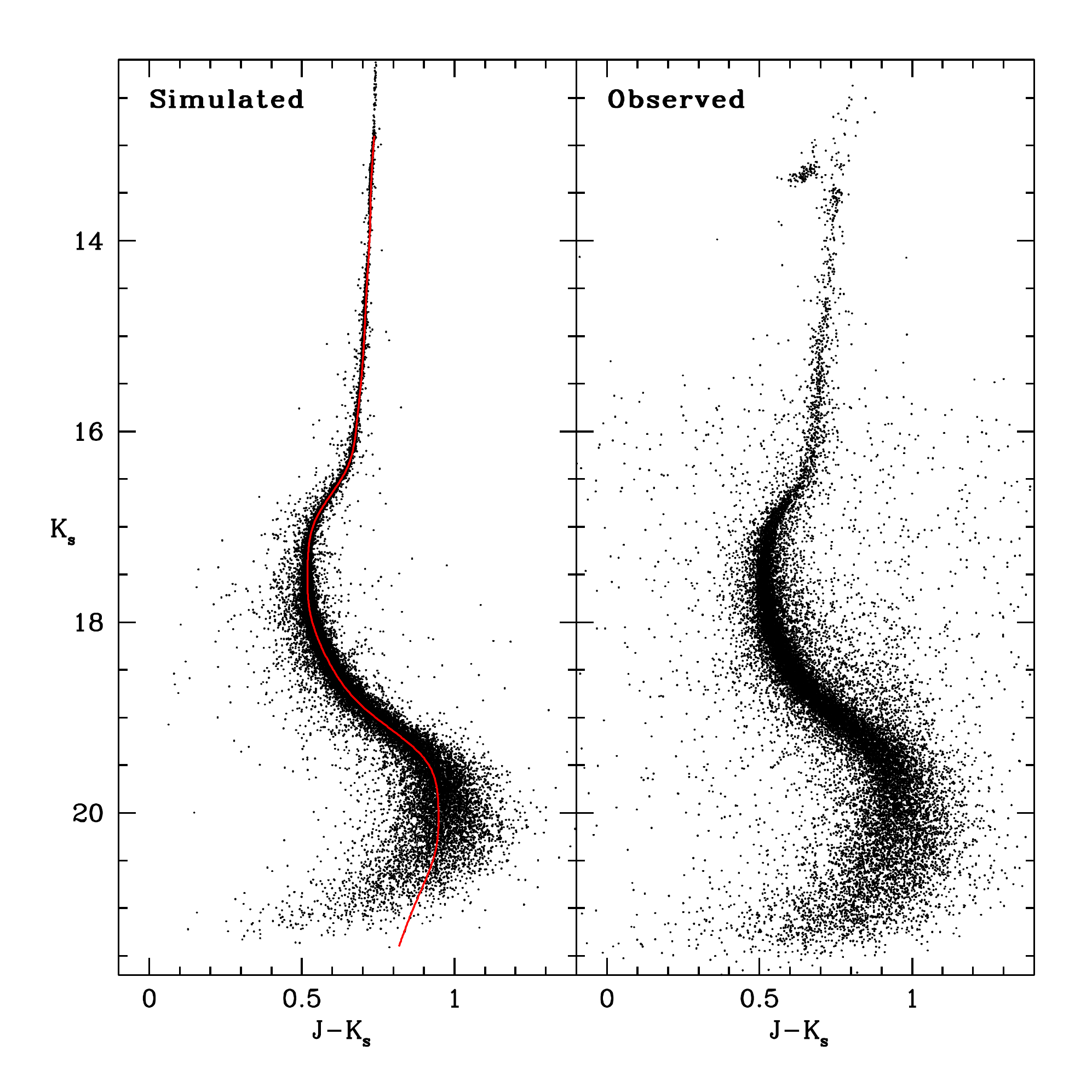}
\caption{Simulated ({\it left} panel) and observed ({\it right} panel)
  ($K_s, J-K_s$) CMD of NGC 6624, for all the stars that survived the 
  selection: -0.2 $\le$ sharpness $\le$ 0.2. As
  clearly visible, the two CMDs turn out to be fully comparable, especially in
  the MS region (at $K_{s}$ $>$ 16). The red line in the left panel represents the MRL of NGC 6624 in the ($K_s, J-K_s$) CMD.}
\label{sim_conf}
\end{figure}   

\begin{figure}
\epsscale{1.}
\plotone{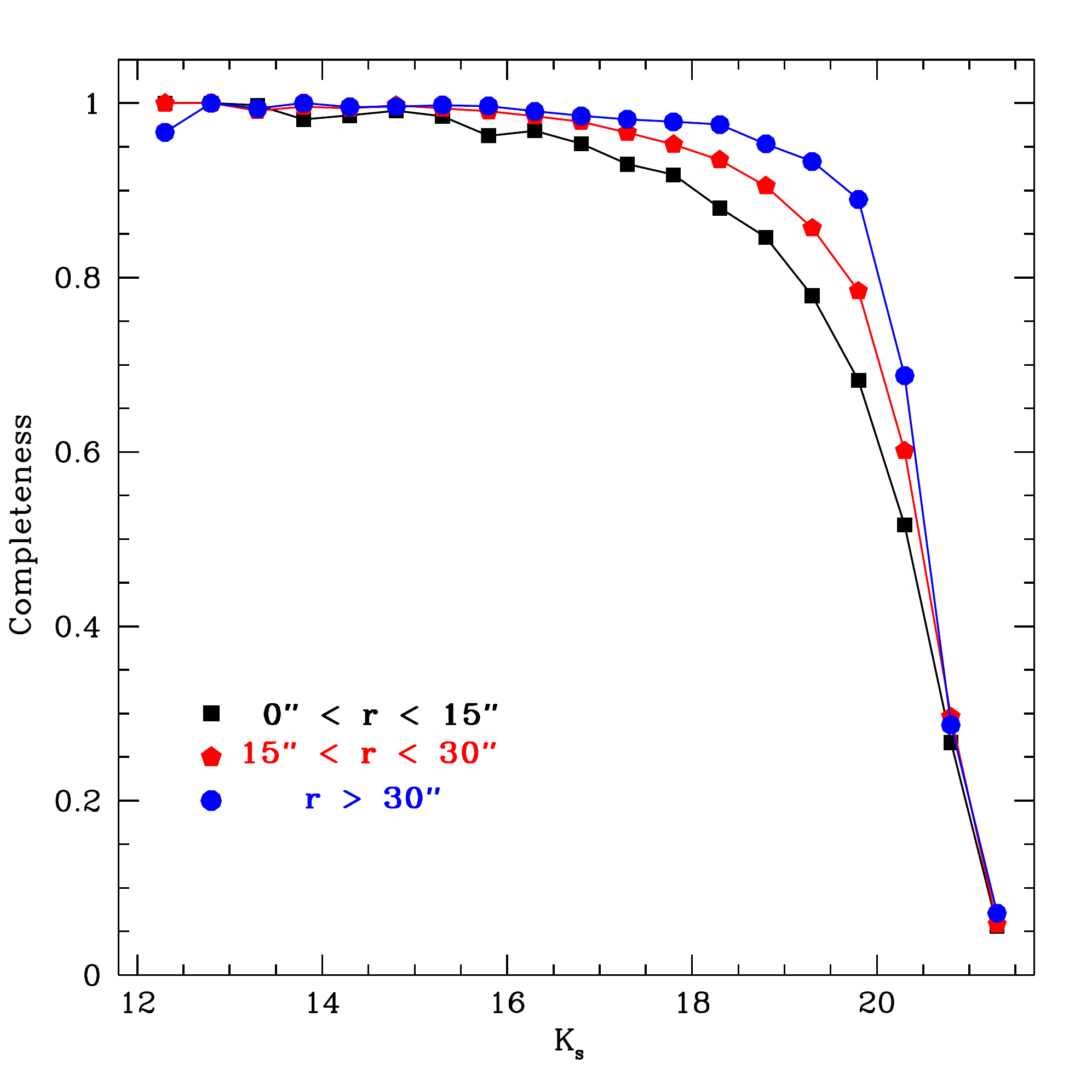}
\caption{Photometric completeness ($\Gamma$) as a function of the
  $K_s$ magnitude for the GEMINI catalog of NGC 6624 in three
  different radial bins (see labels).}
\label{compl}
\end{figure}      

\begin{figure}
\epsscale{1.}
\plotone{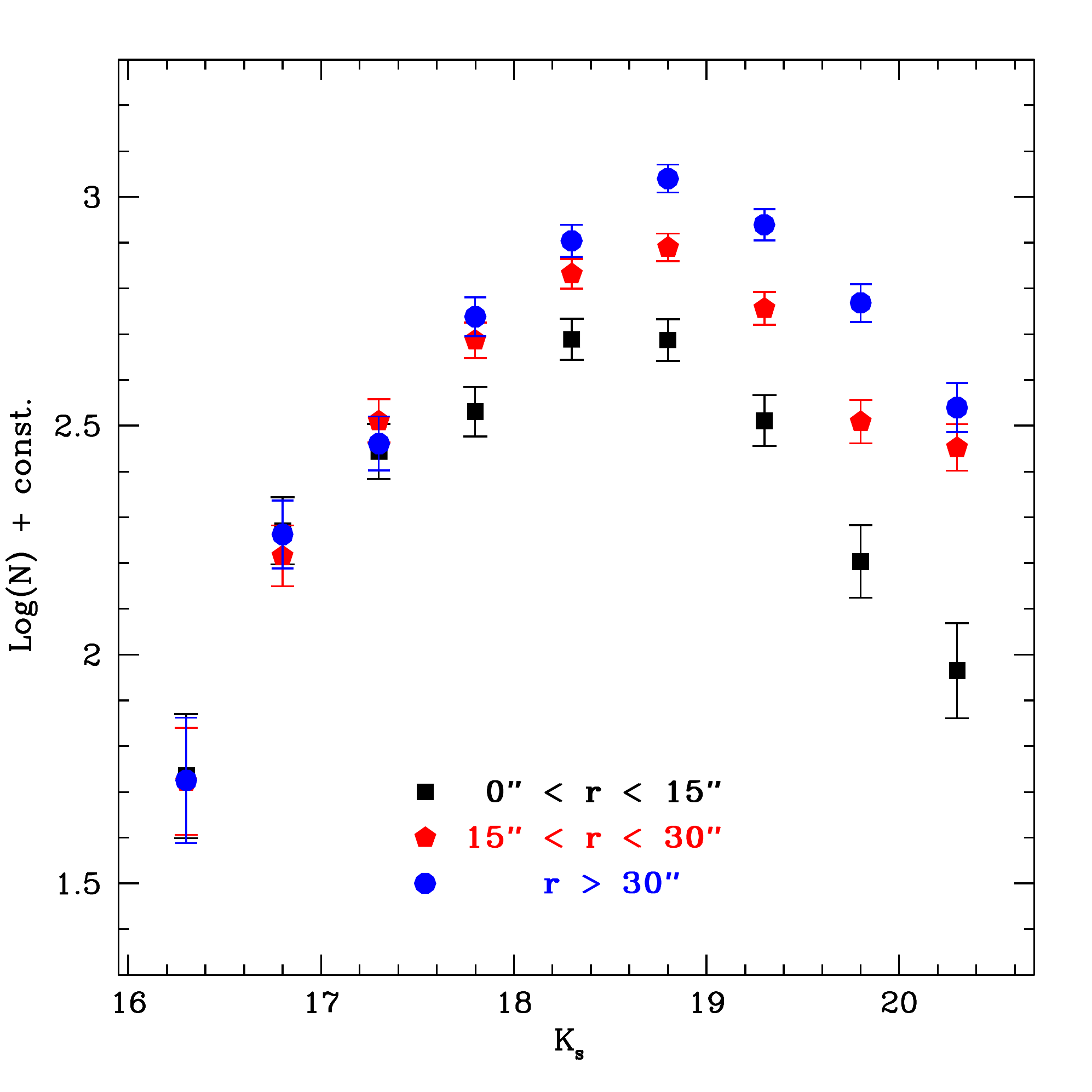}
\caption{NGC 6624 completeness-corrected and field-decontaminated
  MS-LFs in the $K_s$ band obtained from the GEMINI catalog in three
  different radial bins. The LF corresponding to the innermost radial
  bin is used as reference to normalize those at larger radii at the brightest bin.}
\label{LF}
\end{figure} 

\begin{figure}
\epsscale{1.}
\plotone{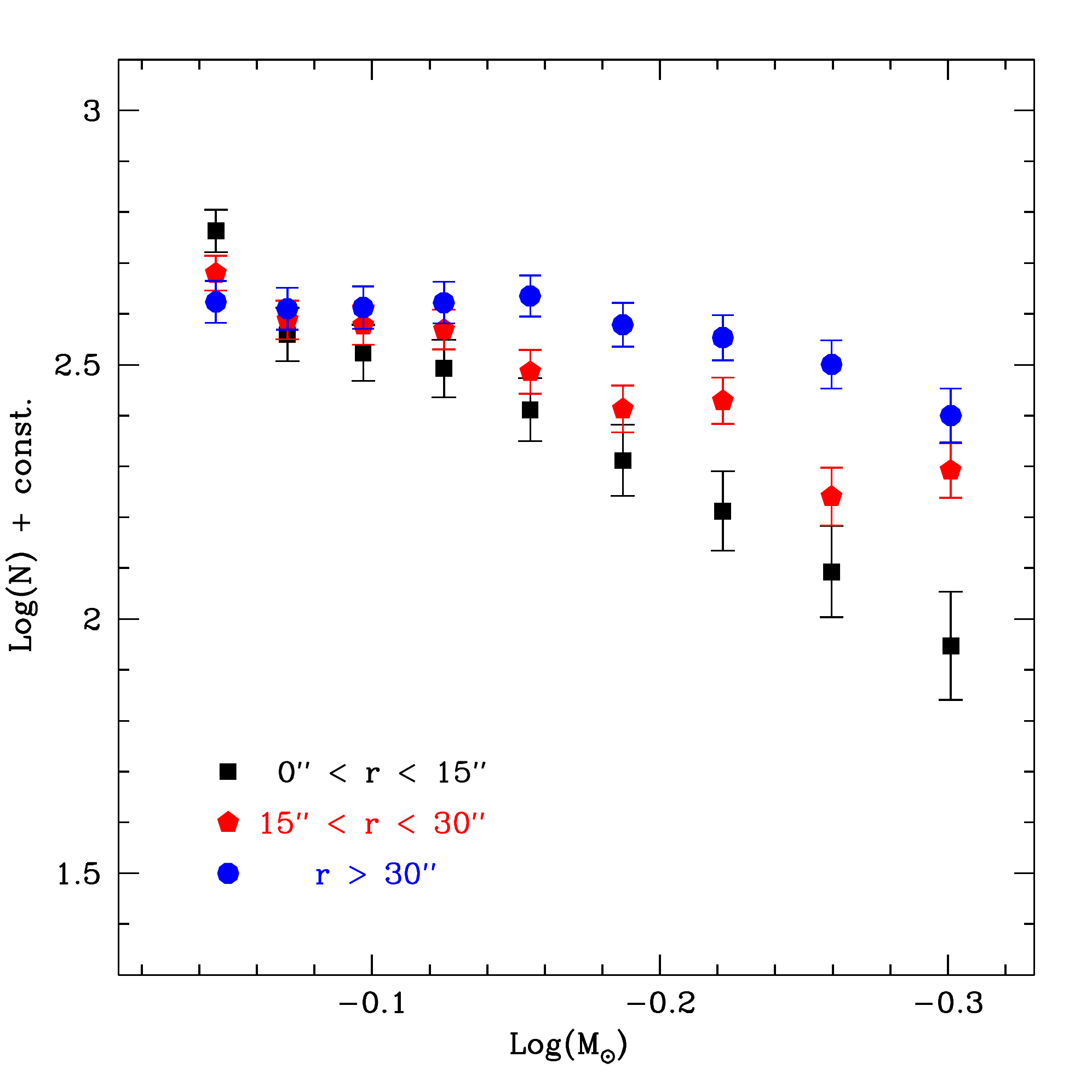}
\caption{MFs derived by using a \texttt{BaSTI} isochrone with [Fe/H] = -0.60
  and $t_{age}$ = 12.0 Gyr (see Section 4.1). Radial bins, symbols and colors are the same as
  in Figure \ref{LF}.}
\label{MF}
\end{figure} 

\begin{figure}[!htb]
\minipage{0.5\textwidth}
  \includegraphics[width=\linewidth,trim={4.5cm 2cm 8cm 3.5cm},clip]{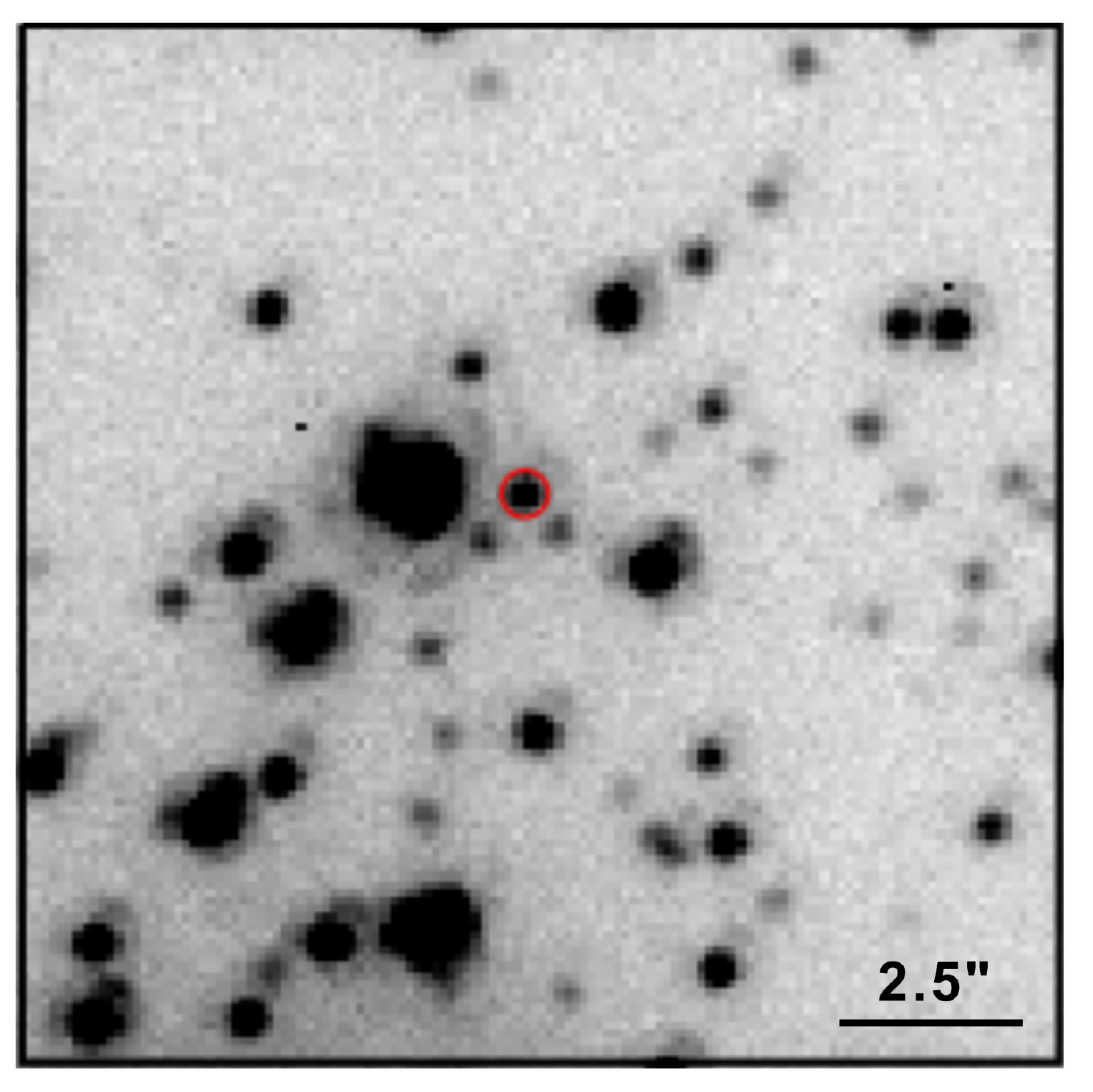}
\endminipage\hfill
\minipage{0.5\textwidth}
  \includegraphics[width=\linewidth,trim={4.5cm 2cm 8cm 3.5cm},clip]{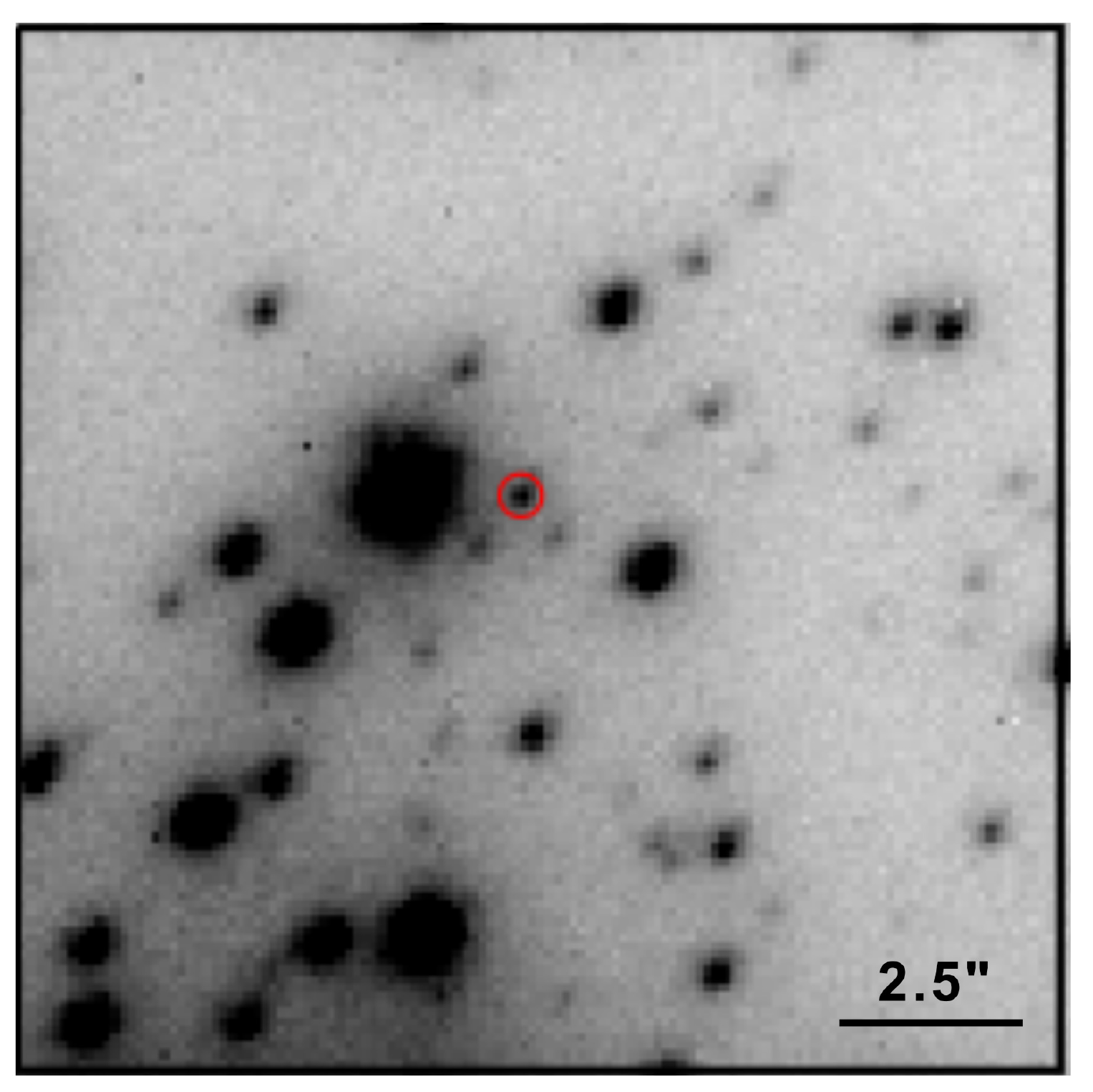}
\endminipage
  \caption{{\it Left panel}: $K_{s}$ image of a region of
    ($2.5\arcsec \times 2.5\arcsec$) of NGC 6624, centered on the
    position of ComStar1, marked with a red circle, as observed by the
    GeMS/GSAOI system. {\it Right panel}: The same, but for the $J$
    band.}
\label{ComStar1}
\end{figure}

\begin{figure}[!htb]
\minipage{0.5\textwidth}
  \includegraphics[width=\linewidth,trim={4.5cm 2cm 8cm 3.5cm},clip]{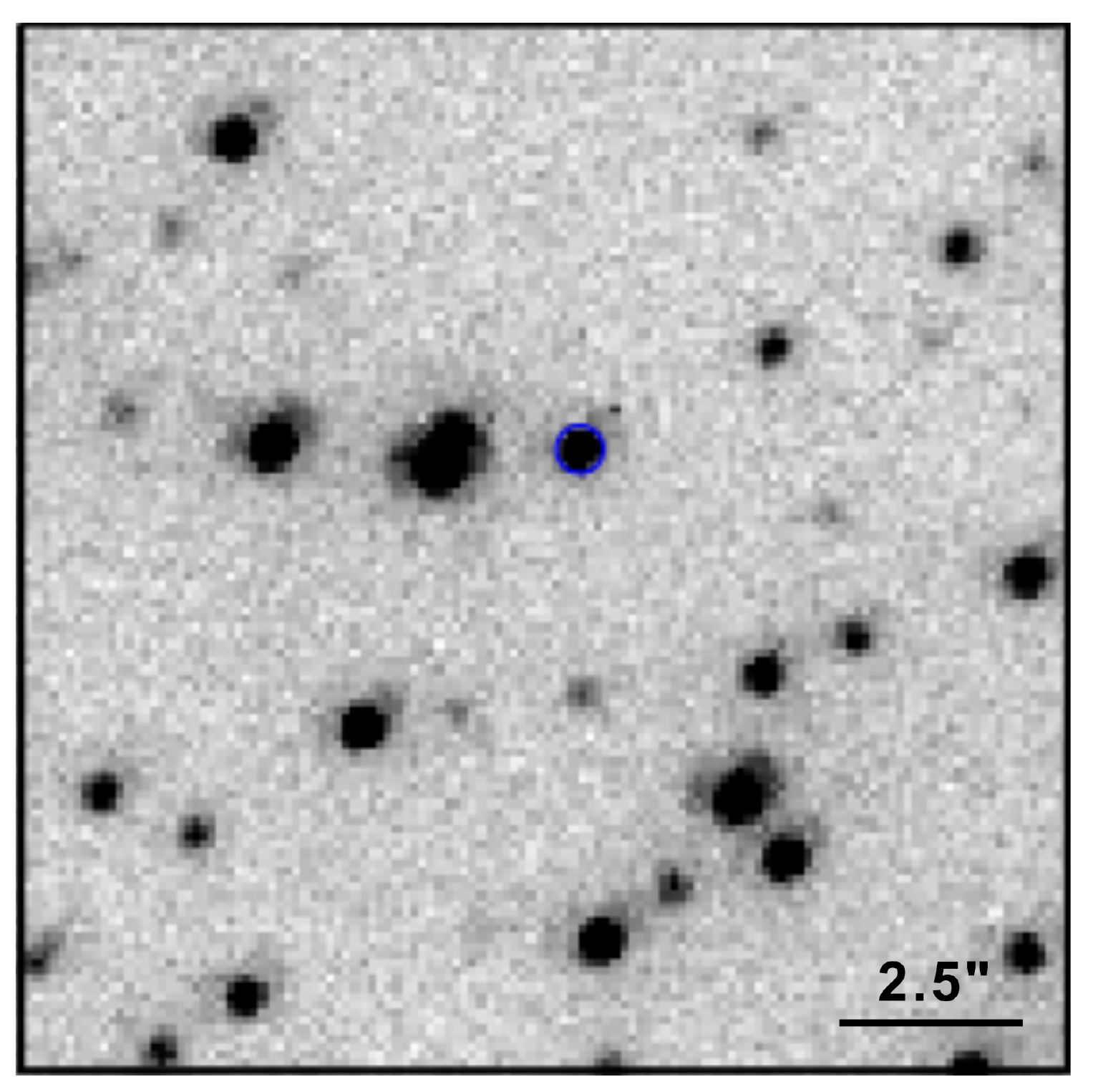}
\endminipage\hfill
\minipage{0.5\textwidth}
  \includegraphics[width=\linewidth,trim={4.5cm 2cm 8cm 3.5cm},clip]{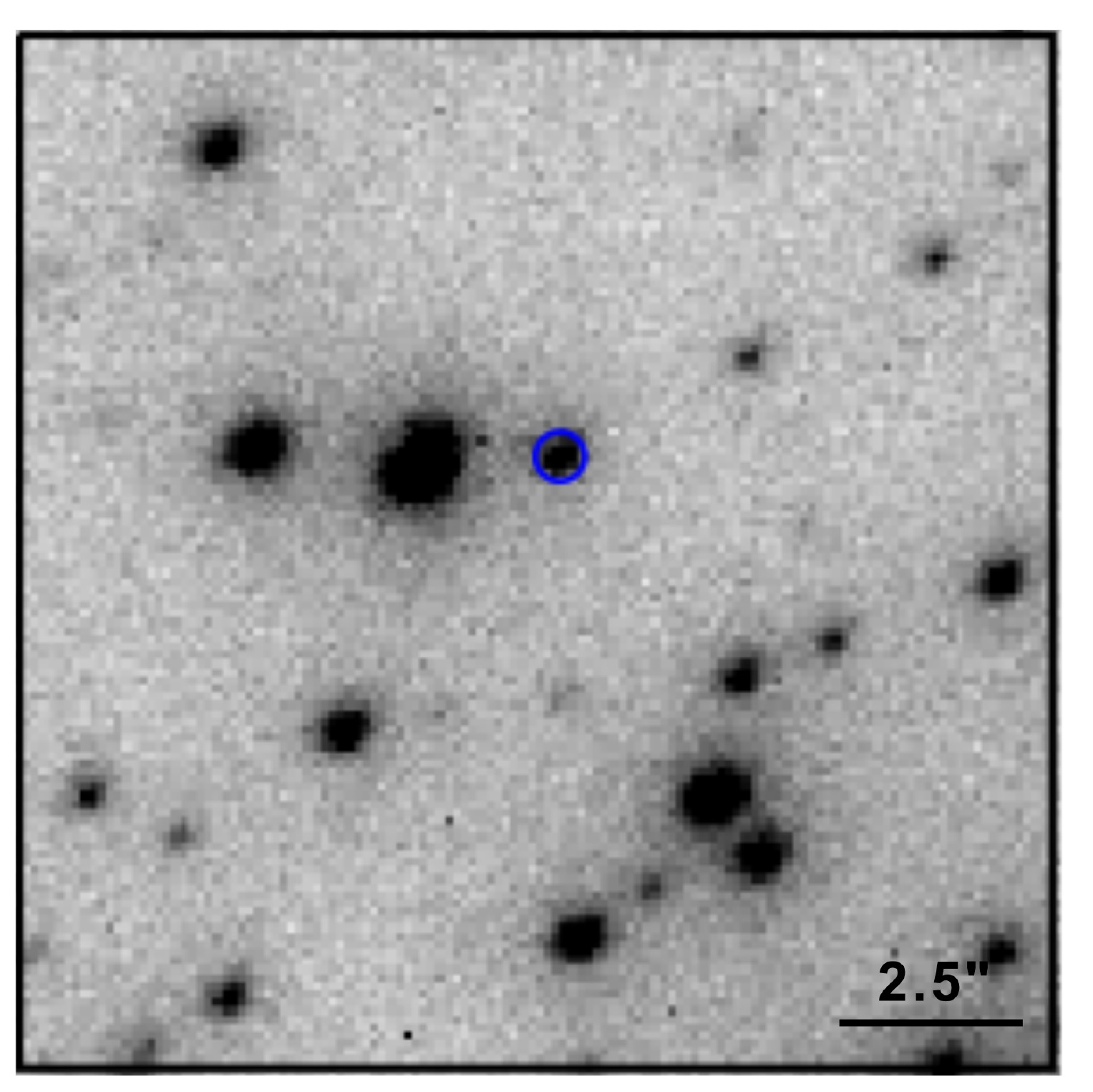}
\endminipage
  \caption{{\it Left panel}: $K_{s}$ image of a region of
    ($2.5\arcsec\; \times 2.5\arcsec$) of NGC 6624, centered on the
    position of StarB, marked with a blue circle, as observed by the
    GeMS/GSAOI system. This star has the same magnitude as ComStar1
    and is adopted for a comparison. {\it Right panel}: The same, but
    in the $J$ band.}
\label{StarB}
\end{figure}

\begin{figure}
\epsscale{1.}
\plotone{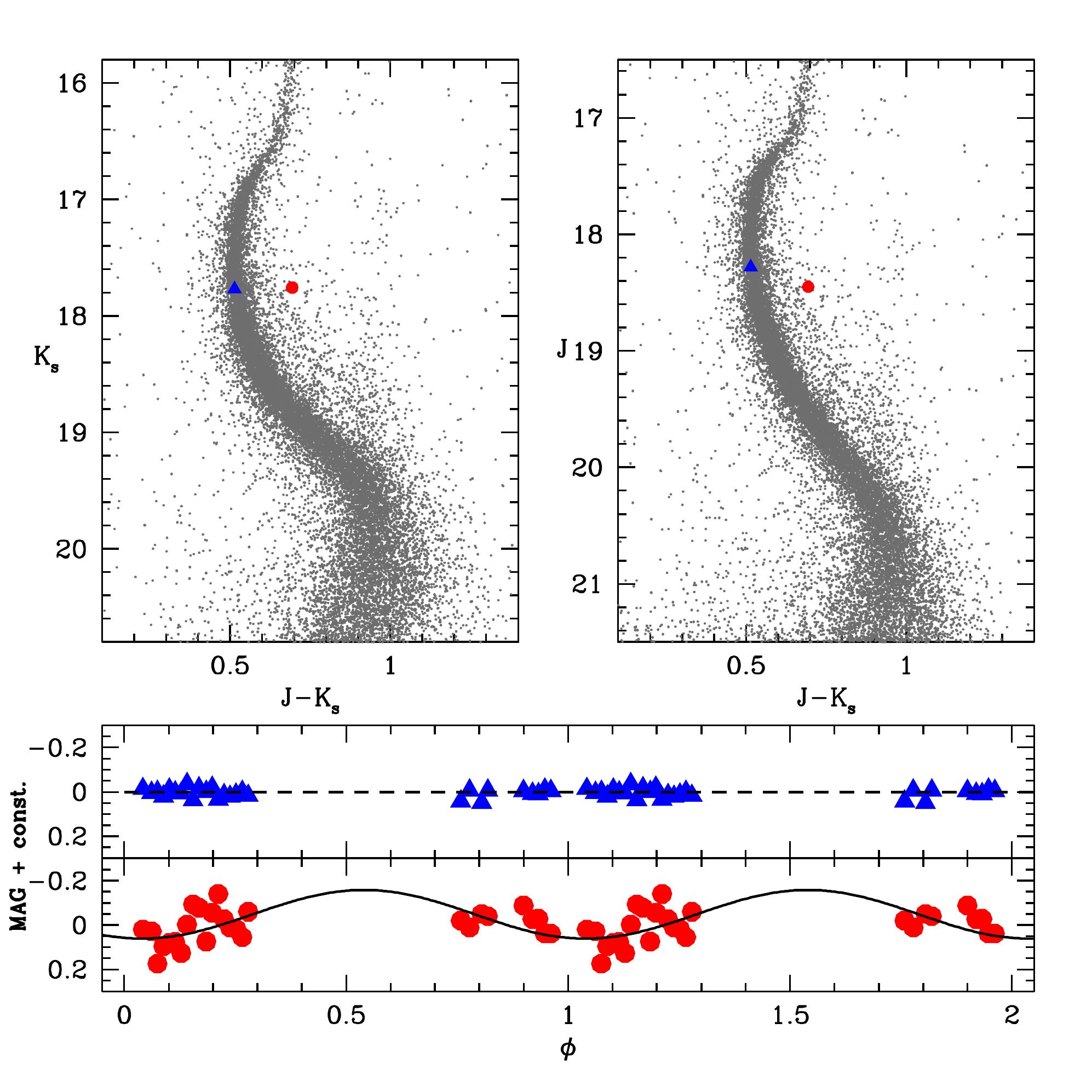}
\caption{{\it Top panels}: Positions in the ($K_s, J-K_s$) and ($J,
  J-K_s$) CMDs of ComStar1 and StarB are highlighted with a red circle
  and a blue triangle, respectively. {\it Bottom panels}: The red
  circles represent the light curve of ComStar1 folded with the
  estimated orbital period $P_{orb}$ $\approx$ 98 min
  \citep{Dal14}. The black line is a sinusoidal function of amplitude
  $\approx$ 0.2. For comparison the blue triangles show the light curve
  of StarB, that has
  the same average $K_{s}$ magnitude of ComStar1 but does not show any
  evidence of flux modulation (dashed black line).}
\label{cmdComStar1}
\end{figure}

\clearpage

\end{document}